\documentclass[aps,pre,twocolumn,10pt,
               superscriptaddress,longbibliography,
               amsmath,amsfonts]{revtex4-2}

\usepackage{mathrsfs,graphicx,comment,xcolor,physics,lipsum}

\usepackage{hyperref}  
\hypersetup{colorlinks,citecolor=red,linkcolor=blue,
            filecolor=black,urlcolor=black}

\begin{document}
\title{Restart and first detection in a lackadaisical quantum walk with flat-band localization}
\author{Debraj Das}
\email{ddas@sissa.it}
\affiliation{SISSA -- International School for Advanced Studies, Via Bonomea 265, 34136 Trieste, Italy}
\affiliation{Istituto dei Sistemi Complessi, Consiglio Nazionale delle Ricerche, via Madonna del Piano 10, 50019 Sesto Fiorentino, Italy}

\begin{abstract}
We study stochastic and sharp restart in a one-dimensional lackadaisical discrete-time quantum walk with self-loop weight $\ell$. In the absence of restart, the dynamics has a flat band responsible for intrinsic localization and two dispersive bands supporting ballistic propagation. We compare two initially localized benchmark states: a flat-band-active  state with finite flat-band overlap and a flat-band-dark state with zero flat-band overlap. For geometric stochastic restart with per-step restart probability $q$, the stationary mean-squared displacement scales as $q^{-2}$ as $q\to0$. In the same limit, the restart-site occupation probability approaches the restart-free intrinsic localized value for the flat-band-active state, whereas for the flat-band-dark state it vanishes as $q\ln(1/q)$. For power-law restart, where $p_m\propto m^{-s}$ is the probability that the waiting time to the next restart is $m$ steps, a normalized stationary site-occupation distribution exists only for $s>2$, while  the stationary absolute spatial moment of order $p$ is finite only for  $s>p+2$. In the regime $1<s\leq2$, at every fixed lattice site, the flat-band-active occupation  converges to the intrinsic flat-band profile, while the flat-band-dark occupation tends to zero. We also consider monitored first detection with sharp restart, in which the walk is reinitialized after a fixed number $r$ of consecutive unsuccessful measurements. For fixed $r$, the mean first-detected-passage time of the flat-band-active state exhibits  a minimum at an intermediate self-loop weight, whereas the flat-band-dark state approaches a ballistic detection limit as $\ell\to\infty$.
\end{abstract}

\maketitle

\section{Introduction}
\label{sec:introduction}

Restart provides a general means of controlling the long-time behavior
of a dynamical process. In stochastic restart, the evolution of a system is
interrupted after random waiting times and the system is reinitialized
in a prescribed state. Even for ordinary diffusion, this intervention
can convert unbounded spreading into a nonequilibrium stationary
distribution and strongly modify first-passage properties~\cite{evans_diffusion_2011,evans_stochastic_2020,
reuveni_optimal_2016,pal_first_2017,das_discrete_2022}.
Renewal theory provides a natural description because each
reinitialization begins a new dynamical interval~\cite{smith_renewal_1958,lapeyre_unified_2024}.
For broadly distributed restart waiting times, however, the
long-time behavior can differ qualitatively from that generated by a
single finite restart scale. In particular, power-law waiting times can produce aging,
and the conditions for the existence of a stationary distribution and
for the convergence of its spatial moments need not coincide~\cite{nagar_diffusion_2016,eule_non-equilibrium_2016,
singh_general_2022,lapeyre_unified_2024}.

Restart has also emerged as a useful tool for controlling quantum
dynamics in recent years~\cite{navascues_resetting_2018,mukherjee_quantum_2018,
kulkarni_generating_2023,dubey_quantum_2023,das_quantum_2022-1, roy_causality_2025, carollo_stochastic_2026, murauer_nonequilibrium_2026, gotta_towers_2026}.
Random reinitialization of closed quantum systems can generate
nonequilibrium states that retain coherent contributions~\cite{mukherjee_quantum_2018,perfetto_designing_2021},
while related protocols have been used to control quantum-jump
statistics in open systems~\cite{perfetto_thermodynamics_2022}.
Quantum walks are a particularly natural setting because coherent
interference, ballistic transport, and spectral localization can all
affect the response to interruption~\cite{nayak_quantum_2000, kempe_quantum_2003, shenvi_quantum_2003, ambainis_quantum_2007, chandrashekar_optimizing_2008, childs_universal_2009, portugal_quantum_2013}. 
Restart has been investigated for
quantum walks on networks, discrete-time unitary dynamics, absorbing
boundaries, quantum-classical transport comparisons, and
continuous-time tight-binding systems~\cite{wald_classical_2021,wald_stochastic_2025,
chelminiak_discrete-time_2025,wojcik_re-examining_2026,dattagupta_stochastic_2022,
acharya_tight-binding_2023}.
A distinct protocol in which only an internal subspace is restarted,
rather than the complete walker-coin state, can instead generate an
effective drifted-diffusion regime~\cite{qiao_emergence_2026}.

A physically different use of restart arises in monitored quantum
first-detection problems, where a target is repeatedly probed by
projective measurements, so every null result modifies the state and
therefore the subsequent evolution. First-detection probabilities
cannot consequently be obtained by simply sampling the unmonitored
position distribution~\cite{dhar_detection_2015,dhar_quantum_2015,
friedman_quantum_2017,thiel_first_2018,das_quantum_2022}.
Under sharp restart, the monitored process is reinitialized after a
prescribed number of unsuccessful detection attempts. Such protocols
can accelerate quantum hitting and reduce the mean
first-detected-passage time~\cite{yin_restart_2023,shukla_accelerated_2025}.
Quantum interference can also produce multiple extrema, staircase
structures, and abrupt changes of the optimal restart threshold~\cite{yin_instability_2024}.
Monitored recurrence in a one-parameter family of three-state quantum
walks has likewise been shown to depend sensitively on both the coin
parameter and the initial coin state~\cite{stefanak_monitored_2023}.
Throughout this work, we use \emph{restart} as the common terminology,
while distinguishing stochastic restart of the unmonitored walk from
sharp restart of the monitored first-detection process.

The underlying dynamics considered here is a one-dimensional
lackadaisical discrete-time quantum walk (LDTQW). Lackadaisical quantum
walks extend coined quantum walks by introducing self-loops~\cite{wong_grover_2015}, 
which can be represented by a single weighted
loop with continuous weight $\ell$~\cite{wong_coined_2017}. On the 
one-dimensional lattice, the weighted
walk is closely related to the three-state Grover walk and its
continuous deformations~\cite{inui_one-dimensional_2005,stefanak_continuous_2012}.
In the absence of restart, its spectrum contains two qualitatively distinct 
dynamical sectors: an
intrinsically localized flat band and dispersive bands supporting
ballistic propagation. Localization, propagation velocities, long-time
distributions, and ballistic spreading in such three-state and
lackadaisical walks have been studied in detail~\cite{wang_one-dimensional_2017,falkner_weak_2014,
stefanak_limit_2014,falcao_universal_2021}.

This coexistence of flat and dispersive bands creates a restart problem
that is qualitatively different from one in which the underlying
dynamics is entirely spreading. 
In a purely dispersive walk, restart would confine every initial 
state in qualitatively the same way, leaving no dynamical 
signature by which an intrinsic localization mechanism could 
be distinguished from confinement generated by the restart itself.
Probability in the flat band can remain localized without 
restart, whereas probability in the
dispersive bands propagates ballistically and can be confined by
repeated interruption. Moreover, the relative population of these
sectors is controlled by the initial coin state. Restart can therefore
act as an operational spectral discriminator: it probes the localized
and propagating components of the same unitary evolution differently
and allows intrinsic localization to be distinguished from confinement
generated by restart.

To exploit this structure, we introduce two benchmark preparations.
The \emph{flat-band-active} state has finite overlap with the flat-band localized
sector, whereas the \emph{flat-band-dark} state has identically zero
flat-band projection and is supported entirely in the dispersive
sector. The latter terminology refers specifically to the absence of
intrinsic flat-band weight and should not be confused with a
detector-dark state or subspace in monitored first-detection theory.
The active-dark comparison therefore provides a controlled way to
separate spectral localization inherited from the restart-free walk
from effects generated by interruption.

We first consider stochastic restart of the complete walker-coin
state. For geometric stochastic restart, the two benchmark states
show sharply different weak-restart signatures. As the per-step
restart probability $q$ tends to zero, the restart-site occupation of
the flat-band-active state approaches its restart-free intrinsic
localized value, whereas for the flat-band-dark state it vanishes as
$q\ln(1/q)$. This nonanalytic dependence is the renewal imprint of the
algebraic $t^{-1}$ return probability of the dispersive walk. At the
same time, restart cuts off ballistic excursions on the scale $q^{-1}$,
producing a ballistic-to-confined crossover and a stationary
mean-squared displacement that diverges as $q^{-2}$ in the
weak-restart limit.

Power-law restart reveals an even sharper distinction between local and
global long-time behavior. For power-law restart, where
$p_m \propto m^{-s}$ is the probability that the next
restart occurs after $m$ steps, a normalized stationary
site-occupation distribution exists only for $s>2$, while for the
benchmark states considered here a finite stationary $p$th absolute
moment requires $s>p+2$. In particular, the stationary mean-squared
displacement is finite only for $s>4$. For $1<s\leq2$, the mean waiting
time diverges and no normalized stationary spatial distribution
exists. Nevertheless, in this same regime, the flat-band-active
occupation at every fixed lattice site converges to the intrinsic
flat-band profile, while the complementary dispersive probability is
transported to increasingly large distances. For the flat-band-dark
preparation the fixed-site occupation instead vanishes. Heavy-tailed
restart therefore separates local spectral memory from global
probabilistic stationarity.

We finally consider monitored first detection under sharp restart.
Here the measurements themselves alter the dynamics, but the same
flat-band-dispersive competition remains visible. The fixed number $r$
of consecutive unsuccessful measurements before restart and the
self-loop weight jointly control the mean first-detected-passage time.
For fixed $r$, the mean first-detected-passage time of the
flat-band-active preparation has a minimum at an intermediate
self-loop weight, reflecting competition between faster dispersive
transport and the increasingly nonpropagating character of the active
preparation at large $\ell$.
The flat-band-dark state avoids this suppression and approaches a
simple large-$\ell$ ballistic detection regime. The self-loop weight is
therefore not merely a transport parameter: it continuously reshapes
the balance between localized and propagating behavior and thereby
modifies the optimal sharp-restart protocol. Stochastic restart and
monitored sharp restart act through different dynamical mechanisms, yet
both expose how the spectral structure of the LDTQW controls its
response to interruption.

The paper is organized as follows.
Section~\ref{sec:LDTQW} introduces the restart-free LDTQW,
derives its flat and dispersive spectral sectors, and defines the
flat-band-active and flat-band-dark benchmark states.
Section~\ref{sec:single_site_restart} develops the discrete renewal
formalism for stochastic restart.
Sections~\ref{sec:geometric_restart} and
\ref{sec:powerlaw_restart} apply this framework to geometric and
power-law restart statistics, respectively.
Section~\ref{sec:sharp_restart} considers monitored first detection
under sharp restart.
Section~\ref{sec:conclusion} summarizes the main results, while
technical derivations are collected in the Appendices.

\section{Restart-free lackadaisical discrete-time quantum walk (LDTQW)}
\label{sec:LDTQW}

We consider the weighted-self-loop formulation of the one-dimensional
LDTQW~\cite{wong_grover_2015, wang_one-dimensional_2017, 
stefanak_limit_2014}, which is equivalent to a continuous
deformation of the three-state Grover walk
~\cite{shenvi_quantum_2003, stefanak_continuous_2012}. Its spectrum contains
a localized flat-band sector coexisting with ballistically propagating
modes~\cite{wang_one-dimensional_2017}. We formulate the model below in
the notation required for the restart problem and isolate its
flat-band contribution explicitly.

\subsection{Model and restart-free propagator}
\label{subsec:LDTQW_model}

The Hilbert space of the walker is
$\mathcal{H}=\mathcal{H}_C\otimes\mathcal{H}_P$, where the 
position space
$\mathcal{H}_P$ is spanned by the orthonormal basis
$\{\ket{n}:n\in\mathbb{Z}\}$ and the three-dimensional coin space
$\mathcal{H}_C$ is spanned by the orthonormal basis 
$\{\ket{-1},\ket{0},\ket{+1}\}$. 
The three coin basis states correspond,
respectively, to a displacement by one lattice site to the left, no
displacement, and a displacement by one lattice site to the right. The
state of the walker at time $t$ is
$\ket{\Psi(t)}=\sum_n\ket{\psi_n(t)}\otimes\ket n$, where 
$\ket{\psi_n(t)}$ denotes the local three-component coin 
vector at site $n$ and reads  
$\ket{\psi_n(t)}=\sum_{c=-1}^{1}\psi_n^{(c)}(t)\ket c$ with 
$\psi_n^{(c)}(t)$ being the corresponding probability
amplitudes.

We introduce the self-loop weight $\ell$ through the normalized
coin state
\begin{align}
    \ket{s_\ell}
    =
    \frac{1}{\sqrt{2+\ell}}
    \left(
        \ket{-1}
        +\sqrt{\ell}\ket{0}
        +\ket{+1}
    \right),
    \label{eq:flat-band-active-sl}
\end{align}
and define the weighted Grover coin as the reflection
$C_\ell \equiv 2\ket{s_\ell}\bra{s_\ell}-\mathbb{I}_3$, where 
$\mathbb{I}_3$ is the identity operator in $\mathcal{H}_C$~\cite{shenvi_quantum_2003}.
In the ordered basis
$\{\ket{-1},\ket{0},\ket{+1}\}$, this gives
\begin{align}
C_\ell
=
\frac{1}{2+\ell}
\begin{pmatrix}
-\ell & 2\sqrt{\ell} & 2 \\
2\sqrt{\ell} & \ell-2 & 2\sqrt{\ell} \\
2 & 2\sqrt{\ell} & -\ell
\end{pmatrix}.
\label{eq:coin_matrix}
\end{align}
For $\ell=1$, Eq.~\eqref{eq:coin_matrix} reduces to the usual
three-state Grover coin~\cite{inui_one-dimensional_2005, shenvi_quantum_2003}. 
We consider $\ell>0$ throughout; the limiting
case $\ell=0$, as shown below, is spectrally degenerate because the dispersive bands
become flat.

The conditional shift operator is
\begin{align}
    S
    =
    \sum_{n=-\infty}^{\infty}
    \sum_{c=-1}^{1}
    \ket{c}\bra{c}
    \otimes
    \ket{n+c}\bra{n}.
    \label{eq:lqw_shift}
\end{align}
Consequently, conditioned on the coin states
$\ket{-1}$, $\ket{0}$, and $\ket{+1}$, the shift $S$ produces a leftward
displacement, no displacement, and a rightward displacement,
respectively. 
Evolution over a single time step is generated by 
$\ket{\Psi(t+1)}= U_\ell \ket{\Psi(t)}$, where 
$U_\ell =S[C_\ell\otimes\mathbb I_P]$ with $\mathbb I_P$ being 
the identity operator in $\mathcal{H}_P$.

Translation invariance allows the walk to be diagonalized in
a continuous quasimomentum Fourier space. 
Setting the lattice spacing to unity, we take the quasimomentum 
$k\in[-\pi,\pi)$, corresponding to the first Brillouin zone, and use
the Fourier convention
$\ket*{\widetilde{\psi}(k,t)}
    =
    \sum_{n=-\infty}^{\infty}
    e^{-ikn}\ket*{\psi_n(t)}$,
with inverse transformation
$\ket*{\psi_n(t)}
=(2\pi)^{-1}\int_{-\pi}^{\pi}
dk\,e^{ikn}\ket*{\widetilde{\psi}(k,t)}$.
In Fourier space, the shift~\eqref{eq:lqw_shift} becomes
$S(k)=\operatorname{diag}(e^{ik},1,e^{-ik})$, and hence  
$U_\ell(k) = S(k) C_\ell$, which reads 
\begin{align}
U_\ell(k) 
=
\frac{1}{2+\ell}
\begin{pmatrix}
-\ell e^{ik} & 2\sqrt{\ell}\,e^{ik} & 2e^{ik} \\
2\sqrt{\ell} & \ell-2 & 2\sqrt{\ell} \\
2e^{-ik} & 2\sqrt{\ell}\,e^{-ik} & -\ell e^{-ik}
\end{pmatrix}.
\label{eq:momentum_matrix}
\end{align}
A single time step evolution in this space is given by
$\ket*{\widetilde{\psi}(k,t+1)}
=U_\ell(k)\ket*{\widetilde{\psi}(k,t)}$, and accordingly, after $t$ time steps  
$\ket*{\widetilde{\psi}(k,t)} = [U_\ell(k)]^t \ket*{\widetilde{\psi}(k,0)}$.

We consider a walker initially localized at site $n_0$ with a normalized coin
state $\ket{\chi_0}$, so that one has 
\begin{align}
    &\ket{\Psi(0)}
    =
    \ket{\chi_0}\otimes\ket{n_0}. \label{eq:lqw_initial_state_psi0}
\end{align}
The corresponding local 
coin vector is
$\ket{\psi_n(0)}=\delta_{n,n_0}\ket{\chi_0}$ with its Fourier transform
$\ket*{\widetilde{\psi}(k,0)}
=e^{-ikn_0}\ket{\chi_0}$.
Applying the inverse Fourier transform, we obtain
\begin{align}
    \ket{\psi_n(t)}
    =
    K_{n-n_0}(t)\ket{\chi_0} ,
    \label{eq:bare_amplitude_direct}
\end{align}
where the restart-free spatial propagator $K_{n-n_0}(t)$ is given by
\begin{align}
    K_{n-n_0}(t)
    \equiv
    \frac{1}{2\pi}
    \int_{-\pi}^{\pi}
    dk\,
    e^{ik(n-n_0)}
    [U_\ell(k)]^t .
    \label{eq:bare_matrix_propagator}
\end{align}
The corresponding site-occupation probability reads
\begin{align}
    P^{(0)}(n,t|n_0,\chi_0)
    = \braket{\psi_n(t)}{\psi_n(t)} =
    \left\|
        K_{n-n_0}(t)\ket{\chi_0}
    \right\|^2 ,
    \label{eq:bare_probability}
\end{align}
where the superscript $(0)$ denotes restart-free evolution.

\subsection{Quasienergy spectrum}
\label{subsec:LDTQW_spectrum}

\begin{figure}[tbp]
    \centering
    \includegraphics[width=1\linewidth]{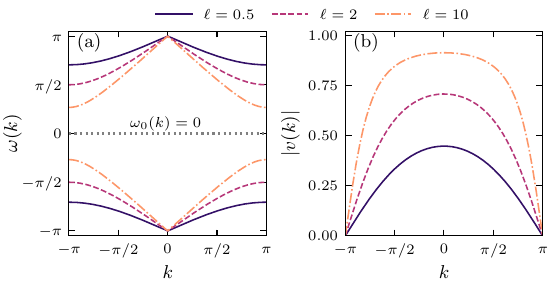}
    \caption{
    (a) Quasienergy spectrum of the restart-free LDTQW for several
    self-loop weights $\ell$. The flat band remains at 
    $\omega_0(k)=0$ (gray dotted line), while the dispersive branches
    $\pm\omega(k)$, obtained from  Eq.~\eqref{eq:dispersion}, become increasingly
    $k$ dependent and span a larger quasienergy range as $\ell$
    increases. 
    (b) Magnitude of the dispersive group
    velocity $|v(k)|$ obtained from Eq.~\eqref{eq:lqw_group_velocity}.  
    }
    \label{fig:dispersion_velocity}
\end{figure}

We next summarize the spectral structure of the 
LDTQW~\cite{wang_one-dimensional_2017}. 
One may write the eigenvalues of the unitary operator $U_\ell(k)$ as
$\lambda_j(k)=\exp({-i\omega_j(k)})$, where the quasienergies
$\omega_j(k)$ for $j\in\{-1,0,+1\}$ are defined modulo $2\pi$. 
The $k$-independent eigenvalue $\lambda_0=1$ corresponds to the flat
band $\omega_0(k)=0$. 
For the remaining two bands, choosing
$\omega_{\pm 1}(k)=\pm\omega(k)$ one obtains
\begin{align}
    \cos\omega(k)
    =
    -\frac{2+\ell\cos k}{2+\ell}.
    \label{eq:dispersion}
\end{align}
The flat band thus produces a time-independent contribution whenever the
initial state has nonzero overlap with its spectral subspace, whereas
the two dispersive bands generate the propagating part of the wave
packet. Their coexistence underlies the simultaneous localization and
ballistic spreading characteristic of the one-dimensional LDTQW
\cite{wang_one-dimensional_2017}.

For the dispersive branches, the magnitude of the group velocities 
$v_\pm(k)\equiv d\omega_\pm(k)/dk$ obtained from  Eq.~\eqref{eq:dispersion} gives 
\begin{align}
    |v(k)|
    &=
    \sqrt{
        \frac{\ell(1+\cos k)}
             {\ell+4+\ell\cos k}
    } ,
    \label{eq:lqw_group_velocity}
\end{align}
whose maximum
$v_{\max}(\ell)=\sqrt{\ell/(\ell+2)}$ increases monotonically
with $\ell$~\cite{wang_one-dimensional_2017,wong_coined_2017,
stefanak_continuous_2012}. 
Figure~\ref{fig:dispersion_velocity}
illustrates how the quasienergy spectrum and the group 
velocity depend on the self-loop weight.

The increase of $v_{\max}(\ell)$ with the self-loop weight $\ell$ contrasts 
with the intuition of a classical lazy random walk. 
Here, the parameter $\ell$ is not a waiting
probability but a coherent weight entering the Grover reflection. 
In the limit $\ell\rightarrow0$, the coin exchanges the left- and
right-moving states, leading to repeated reversals of the propagating
components; at $\ell=0$ the dispersive bands become flat. 
By contrast, for $\ell\rightarrow\infty$,
$C_\ell\rightarrow\operatorname{diag}(-1,1,-1)$, so the moving coin
states become asymptotically decoupled and propagate persistently with
velocities approaching $-1$ and $+1$. Thus a larger self-loop weight
increases the propagation speed of the dispersive sector rather than
acting as a stochastic slowing mechanism.
Faster dispersive
propagation, however, does not necessarily imply stronger overall
spreading: the spectral weight carried by the dispersive bands also
depends on $\ell$ and on the initial coin state. This
spectral--kinematic competition will become important below.

\subsection{Intrinsic localization}
\label{subsec:LDTQW_intrinsic_localization}

Localization in three-state quantum walks is well established: it
arises in the three-state Grover walk~\cite{inui_one-dimensional_2005}
and persists under continuous Grover-type coin
deformations~\cite{stefanak_continuous_2012,
wang_one-dimensional_2017}, and has recently been 
characterized rigorously for general coin matrices~\cite{kiumi_spectral_2026}.
Let us isolate the flat-band and dispersive contributions
directly at the propagator level. Since $\lambda_0=1$, the
quasimomentum evolution operator admits the spectral
decomposition
\begin{align}
    [U_\ell(k)]^t
    =
    \mathcal{P}_0(k)
    +
    e^{-i\omega(k)t}\mathcal{P}_+(k)
    +
    e^{i\omega(k)t}\mathcal{P}_-(k) ,
    \label{eq:lqw_spectral_decomposition}
\end{align}
where $\mathcal{P}_{0}(k)$ and $\mathcal{P}_{\pm}(k)$ denote 
the spectral projectors associated with the flat 
and dispersive bands, respectively.
Substituting Eq.~\eqref{eq:lqw_spectral_decomposition} in 
Eq.~\eqref{eq:bare_matrix_propagator} yields 
\begin{align}
    K_{n-n_0}(t)
    =
    K_{n-n_0}^{\mathrm{fb}}
    +
    K_{n-n_0}^{\mathrm{disp}}(t),
    \label{eq:lqw_prop_decomposition}
\end{align}
where
\begin{align}
    &K_{n-n_0}^{\mathrm{fb}}
    \equiv
    \int_{-\pi}^{\pi}
    \frac{dk}{2 \pi}\,
    e^{ik(n-n_0)}\mathcal{P}_0(k) , \label{eq:lqw_localized_kernel0} \\
    &K_{n-n_0}^{\mathrm{disp}}(t) 
    \equiv
    \int_{-\pi}^{\pi}
    \frac{dk}{2\pi} \,
    e^{ik(n-n_0)} \nonumber \\
    &\hskip60pt \times \left[
        e^{-i\omega(k)t}\mathcal{P}_+(k)
        +
        e^{i\omega(k)t}\mathcal{P}_-(k)
    \right].
    \label{eq:lqw_dispersive_kernel}
\end{align}
Here, the quantity $K_{n-n_0}^{\mathrm{fb}}$ contains the 
time-independent flat-band contribution, 
whereas $K_{n-n_0}^{\mathrm{disp}}(t)$ contains the
time-dependent contribution from the dispersive bands. 
Since this decomposition~\eqref{eq:lqw_prop_decomposition} is at the 
amplitude level (see Eq.~\eqref{eq:bare_amplitude_direct}), the 
localized and dispersive sectors can interfere 
at finite times.

To isolate the flat-band contribution, let us consider a 
normalized flat-band eigenvector $\ket{v_0(k)}$ of $U_\ell(k)$:
\begin{align}
    \ket{v_0(k)}
    =
    \frac{1}{\sqrt{2+\ell\cos^2(k/2)}}
    \begin{pmatrix}
        e^{ik/2}\\[1mm]
        \sqrt{\ell}\cos(k/2)\\[1mm]
        e^{-ik/2}
    \end{pmatrix} .
    \label{eq:lqw_flat_eigenvector}
\end{align}
With $\mathcal P_0(k)=\ket{v_0(k)}\bra{v_0(k)}$, the flat-band 
kernel $K_{n-n_0}^{\mathrm{fb}}$ in Eq.~\eqref{eq:lqw_localized_kernel0} 
can be evaluated explicitly.
Its matrix elements are linear combinations of the Fourier
coefficients
(see Appendix~\ref{app:Gm_integral})
\begin{align}
    G_{n-n_0}(\ell)
    &\equiv
    \frac{[-\eta(\ell)]^{|n-n_0|}}
         {2\sqrt{2(\ell+2)}}, 
    \label{eq:lqw_Gm} 
\end{align}
where we have 
\begin{align}
    \eta(\ell)
    &\equiv
    \frac{\sqrt{\ell+2}-\sqrt{2}}
         {\sqrt{\ell+2}+\sqrt{2}} .
    \label{eq:lqw_eta}
\end{align}
Since $0<\eta(\ell)<1$ for $\ell>0$, the coefficients $G_{n-n_0}(\ell)$
decay exponentially with $|n-n_0|$, producing the
intrinsic flat-band localization, a mechanism closely 
related to flat-band-induced localization in engineered 
lattice systems~\cite{leykam_artificial_2018}.

For a walker initially localized at $n_0$, the intrinsic localized
component depends on the coin state $\ket{\chi_0}$. To keep track of the
flat-band and dispersive sectors separately, we write
$\ket{\psi_n(t)} =
\ket*{\psi_n^{\mathrm{fb}}}
+
\ket*{\psi_n^{\mathrm{disp}}(t)}$
where we have 
\begin{align}
\ket{\psi_n^{\mathrm{fb}}}
\equiv
K_{n-n_0}^{\mathrm{fb}}\ket{\chi_0}, \,\,\,\,
\ket{\psi_n^{\mathrm{disp}}(t)}
\equiv
K_{n-n_0}^{\mathrm{disp}}(t)\ket{\chi_0}.
\label{eq:psi_n_decomp}
\end{align}
Accordingly, the site-occupation probability may be decomposed as
\begin{align}
P^{(0)}(n,t|n_0,\chi_0)
&=
P_{\mathrm{fb}}^{(0)}(n|n_0,\chi_0)
+
P_{\mathrm{disp}}^{(0)}(n,t|n_0,\chi_0) \nonumber \\
&+
P_{\mathrm{int}}^{(0)}(n,t|n_0,\chi_0),
\label{eq:lqw_probability_loc_disp_int}
\end{align}
where the flat-band contribution reads 
\begin{align}
P_{\mathrm{fb}}^{(0)}(n|n_0,\chi_0) \equiv
\braket*{ \psi_n^{\mathrm{fb}} }{ \psi_n^{\mathrm{fb}} }
=
\left\|
        K_{n-n_0}^{\mathrm{fb}}\ket{\chi_0}
    \right\|^2
    \label{eq:lqw_localized_profile},
\end{align}
along with the dispersive and interference contributions given by 
$P_{\mathrm{disp}}^{(0)}(n,t|n_0,\chi_0) \equiv
\braket*{ \psi_n^{\mathrm{disp}}(t) }{ \psi_n^{\mathrm{disp}}(t) }$, 
and 
$P_{\mathrm{int}}^{(0)}(n,t|n_0,\chi_0) \equiv
2 \,\mathrm{Re}
\braket*{ \psi_n^{\mathrm{fb}} }{ \psi_n^{\mathrm{disp}}(t) }$, respectively.
Thus, the quantity $P_{\mathrm{fb}}^{(0)}(n|n_0,\chi_0)$ 
isolates the \emph{intrinsic localized component} of the
restart-free walk. 

The same decomposition is particularly transparent in 
quasimomentum space. One may define the
$k$-resolved flat-band and dispersive spectral weights as 
\begin{equation}
\begin{aligned}
w_{\mathrm{fb}}(k,\chi_0)
&\equiv
\bra{\chi_0}
\mathcal P_0(k)
\ket{\chi_0},
\label{eq:lqw_k_resolved_weights} \\[0.5ex]
w_{\mathrm{disp}}(k,\chi_0)
&\equiv
\bra{\chi_0}
\mathcal P_+(k)+\mathcal P_-(k)
\ket{\chi_0} ,
\end{aligned}
\end{equation}
which satisfy
$ w_{\mathrm{fb}}(k,\chi_0) + w_{\mathrm{disp}}(k,\chi_0) = 1$.
The total flat-band weight reads
\begin{align}
W_{\mathrm{fb}}(\ell,\chi_0)
&\equiv
\sum_n P_{\mathrm{fb}}^{(0)}(n|n_0,\chi_0)
=
\frac{1}{2\pi}
\int_{-\pi}^{\pi}
dk\,w_{\mathrm{fb}}(k,\chi_0) \nonumber \\
&=
\bra{\chi_0}K_0^{\mathrm{fb}}\ket{\chi_0},
\label{eq:W_loc_spectral}
\end{align}
where we have used Parseval's identity and $\mathcal P_0^2=\mathcal P_0$. 
Likewise, the total dispersive spectral weight is given by 
$W_{\mathrm{disp}}(\ell,\chi_0) \equiv
\sum_{n=-\infty}^{\infty}
P_{\mathrm{disp}}^{(0)}(n,t|n_0,\chi_0)
=(2\pi)^{-1}\int_{-\pi}^{\pi}dk\,
w_{\mathrm{disp}}(k,\chi_0)$, and hence
$W_{\mathrm{fb}}+W_{\mathrm{disp}}=1$.

It is important to distinguish the spectral weights
$W_{\mathrm{fb}}$ and $W_{\mathrm{disp}}$ from the 
probability contributions in
Eq.~\eqref{eq:lqw_probability_loc_disp_int}. The former are the total
norms of two mutually orthogonal spectral sectors, whereas the latter
interfere locally in position space. 
For any fixed displacement $n-n_0$, the oscillatory dispersive
contribution dephases as $t\to\infty$, so that
$K_{n-n_0}^{\mathrm{disp}}(t)\ket{\chi_0}\to0$. Hence the restart-free
occupation obeys the pointwise limit
\begin{align}
\lim_{t\to\infty}
P^{(0)}(n,t|n_0,\chi_0)
&=
P_{\mathrm{fb}}^{(0)}(n|n_0,\chi_0).
\label{eq:lqw_pointwise_localized_limit}
\end{align}
This convergence is at fixed $n$: the dispersive spectral weight
generally remains finite but is transported to ballistic distances
$|n-n_0|=O(t)$.

\subsection{Flat-band coin states}
\label{subsec:LDTQW_benchmark_states}

The dependence of $W_{\mathrm{fb}}$ on the coin state becomes
transparent by introducing the states 
$\ket{\chi_{\mathrm A}} \equiv
{(\ket{-1}-\ket{+1})}/{\sqrt{2}}$ and 
\begin{align}
    \ket{\chi_\perp}
    &\equiv
    \frac{
        \sqrt{\ell}\ket{-1}
        -2\ket{0}
        +\sqrt{\ell}\ket{+1}
    }{
        \sqrt{2(\ell+2)}
    } . 
    \label{eq:lqw_dark_state}
\end{align}
The states $\ket{s_\ell}$, $\ket{\chi_\perp}$, and 
$\ket{\chi_{\mathrm A}}$ form the coin
eigenstate basis~\cite{stefanak_limit_2014,falcao_universal_2021}, 
where the zero-displacement flat-band kernel $K_0^{\mathrm{fb}}$ 
becomes diagonal. In particular, one has
\begin{align}
    K_0^{\mathrm{fb}}
    =
    w_{\mathrm{act}}(\ell)
    \ket{s_\ell}\bra{s_\ell}
    +
    w_{\mathrm A}(\ell)
    \ket{\chi_{\mathrm A}}\bra{\chi_{\mathrm A}},
    \label{eq:lqw_K0_spectral}
\end{align}
with
\begin{align}
    w_{\mathrm{act}}(\ell)
    &=
    \frac{\sqrt{\ell+2}}
    {\sqrt{\ell+2}+\sqrt{2}},
    &
    w_{\mathrm A}(\ell)
    &=
    \frac{\sqrt{2}}
    {\sqrt{\ell+2}+\sqrt{2}};
    \label{eq:lqw_flat_eigenvalues}
\end{align}
the eigenvalue associated with $\ket{\chi_\perp}$ is zero. 
For an arbitrary normalized coin state $\ket{\chi_0}$, 
Eqs.~\eqref{eq:W_loc_spectral} and~\eqref{eq:lqw_K0_spectral} yield
\begin{align}
    W_{\mathrm{fb}}(\ell,\chi_0)
    &=
    w_{\mathrm{act}}(\ell)
    \left|\braket{s_\ell}{\chi_0}\right|^2 
    +
    w_{\mathrm A}(\ell)
    \left|\braket{\chi_{\mathrm A}}{\chi_0}\right|^2.
    \label{eq:lqw_flat_weight_general}
\end{align}

For $\ell>0$, we have $w_{\mathrm{act}}(\ell)>w_{\mathrm A}(\ell)$ 
and consequently, for an initially localized walker, 
$\ket{s_\ell}$ maximizes the total flat-band weight among normalized
coin states and yields 
$W_{\mathrm{fb}}(\ell,s_\ell)=w_{\mathrm{act}}(\ell)$.
We therefore use $\ket{s_\ell}$ as a \emph{flat-band-active}
benchmark. Its restart-free evolution contains both a time-independent
localized component and a dispersive propagating component.
The spatial form of the intrinsic localized component can also be
obtained explicitly. Substituting the coefficients
$G_m(\ell)$ from Eq.~\eqref{eq:lqw_Gm} into the flat-band kernel
\eqref{eq:lqw_localized_kernel} and acting on $\ket{s_\ell}$ gives
\begin{align}
    P_{\mathrm{fb}}^{(0)}(n|n_0,s_\ell)
    =
    \begin{cases}
    \displaystyle
    \frac{(1+\eta)^2}{4},
    & n=n_0,
    \\[1.5ex]
    \displaystyle
    \frac{(1-\eta^2)^2}{8}\,
    \eta^{\,2(|n-n_0|-1)},
    & n\neq n_0.
    \end{cases}
    \label{eq:lqw_active_localized_profile}
\end{align}
Thus, the intrinsic component is  symmetric about the initial site $n_0$ and
decays exponentially away from it. 
Summing 
Eq.~\eqref{eq:lqw_active_localized_profile} over the lattice gives
$W_{\mathrm{fb}}(\ell,s_\ell)=w_{\mathrm{act}}(\ell)$, consistently with
Eq.~\eqref{eq:lqw_flat_weight_general}.
Since the contribution of the dispersive modes to the return 
probability $P^{(0)}(n_0,t|n_0,s_\ell)$ vanishes
at long times, one obtains
\begin{align}
    \lim_{t\to\infty}
    P^{(0)}(n_0,t|n_0,s_\ell)
    = P_{\mathrm{fb}}^{(0)}(n_0|n_0,s_\ell) =
    w_{\mathrm{act}}^2(\ell).
    \label{eq:lqw_active_return}
\end{align}

At the opposite extreme, the state $\ket{\chi_\perp}$ is orthogonal to the
flat-band eigenvector for every quasimomentum,
$\braket{v_0(k)}{\chi_\perp}=0$. Hence
$\mathcal P_0(k)\ket{\chi_\perp}=0$ for all $k$, which implies
$K_{n-n_0}^{\mathrm{fb}}\ket{\chi_\perp}=0$,
$P_{\mathrm{fb}}^{(0)}(n|n_0,\chi_\perp)=0$, and 
$W_{\mathrm{fb}}(\ell,\chi_\perp)=0$.
We therefore refer to $\ket{\chi_\perp}$ as \emph{flat-band-dark}, and 
its restart-free
evolution is generated entirely by the two dispersive bands.

For $\ell>0$, a stationary-phase analysis of the dispersive
contribution gives the long-time return probability
(see Appendix~\ref{app:dark_return}) 
\begin{align}
    P^{(0)}(n_0,t|n_0,\chi_\perp)
    \sim
    \frac{1}{\pi}
    \sqrt{\frac{2}{\ell}}\,
    \frac{1}{t},
    \qquad t\to\infty.
    \label{eq:lqw_dark_return}
\end{align}
Thus, the flat-band-dark state can return to the initial site at finite
times, but its return probability decays algebraically and no
persistent localized component survives.
This result agrees with the known $t^{-1}$ return decay of the
detrapping state~\cite{falcao_universal_2021}.

\begin{figure}[tbp]
    \centering
    \includegraphics[width=1\linewidth]{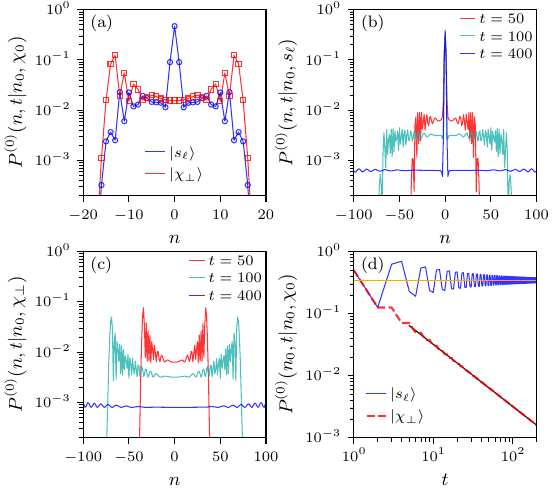}
    \caption{Restart-free dynamics for a walker initially localized at $n_0=0$
    with $\ell=2$.
    (a) Site-occupation probability at $t=20$ for the flat-band-active state
    $\ket{s_\ell}$ (blue) and the flat-band-dark state
    $\ket{\chi_\perp}$ (red). Solid curves show the analytical 
    result~\eqref{eq:bare_probability} and
    open symbols are obtained from numerical simulation.
    (b) and (c) Site-occupation profiles at $t=50$, $100$, and $400$ for
    $\ket{s_\ell}$ and $\ket{\chi_\perp}$, respectively. Lines are obtained 
    from numerical simulation. 
    (d) Return probability at the initial site. The horizontal orange line
    marks $w_{\mathrm{act}}^2$ from Eq.~\eqref{eq:lqw_active_return}, and
    the black line shows the asymptotic result in
    Eq.~\eqref{eq:lqw_dark_return}. The blue solid line and the dashed red 
    line are obtained from numerical simulation.
    }
    \label{fig:restart_free_benchmarks}
\end{figure}

The contrasting restart-free dynamics of the two benchmark states is
shown in Fig.~\ref{fig:restart_free_benchmarks} for $\ell=2$ and
$n_0=0$. Figure~\ref{fig:restart_free_benchmarks}(a) confirms the
analytical site-occupation probability against direct real-space
evolution for both preparations. Their different spectral content
becomes increasingly apparent at longer times. For the flat-band-active
state $\ket{s_\ell}$, Fig.~\ref{fig:restart_free_benchmarks}(b) shows
a persistent central core together with a dispersive component that
propagates ballistically away from the initial site. The central part
is the intrinsic flat-band contribution derived above and remains
localized while the dispersive weight moves to distances of order
$t$. By contrast, the flat-band-dark state $\ket{\chi_\perp}$ in
Fig.~\ref{fig:restart_free_benchmarks}(c) has no flat-band component.
Its entire probability is carried by the dispersive bands, so the
weight moves progressively outward and the occupation of any fixed
region around $n_0$ decreases with time.

The distinction is particularly clear in the return probability in
Fig.~\ref{fig:restart_free_benchmarks}(d). For the flat-band-active
state, the dispersive correction decays and the return probability
approaches the nonzero intrinsic value
$w_{\mathrm{act}}^2(\ell)$ given by
Eq.~\eqref{eq:lqw_active_return}. For the flat-band-dark state, no
persistent local component survives, and the return probability
vanishes algebraically according to
Eq.~\eqref{eq:lqw_dark_return}. The two preparations therefore provide
restart-free reference dynamics with and without intrinsic flat-band
localization, which will allow us to distinguish this mechanism from
restart-induced confinement below.

\subsection{Restart-free mean-squared displacement}

The global spreading of the restart-free walk is characterized by the
mean-squared displacement (MSD) from the initial site, i.e., 
\begin{align}
    M_2^{(0)}(t) \equiv 
    \sum_{n=-\infty}^{\infty}
    (n-n_0)^2
    P^{(0)}(n,t|n_0,\chi_0).
    \label{eq:comp_second_moments_r0}
\end{align}
At long times the spreading is ballistic; one obtains (see
Appendix~\ref{app:restart_free_msd_ballistic}) 
\begin{align}
    M_2^{(0)}(t)
    \sim
    \mathcal B(\ell,\chi_0)t^2,
    \qquad
    t\to\infty ,
    \label{eq:comp_bare_ballistic_r0}
\end{align}
where the coefficient is
\begin{align}
    \mathcal B(\ell,\chi_0)
    &\equiv
    \frac{1}{2\pi}
    \int_{-\pi}^{\pi}
    dk\,
    v^2(k)
    \left[
        1-
        \bra{\chi_0}
        \mathcal P_0(k)
        \ket{\chi_0}
    \right] \nonumber\\
    &=
    \frac{1}{2\pi}
    \int_{-\pi}^{\pi}
    dk\,
    v^2(k)\,
    w_{\mathrm{disp}}(k , \chi_0),
    \label{eq:comp_ballistic_coeff}
\end{align}
Equation~\eqref{eq:comp_ballistic_coeff} makes explicit that the
ballistic MSD is jointly controlled by the propagation speed of the
dispersive modes and by the spectral weight with which these modes are
populated. Each quasimomentum contributes in proportion to
$v^2(k)w_{\mathrm{disp}}(k,\chi_0)$, so a large group velocity alone
does not necessarily imply a large ballistic prefactor. The flat band
itself does not contribute to the leading $t^2$ growth, since its
quasienergy is independent of $k$ and its group velocity therefore
vanishes. 

For the flat-band-active and flat-band-dark preparations introduced in
Sec.~\ref{subsec:LDTQW_benchmark_states},
Eq.~\eqref{eq:comp_ballistic_coeff} gives
(see Appendix~\ref{app:restart_free_msd_ballistic})
\begin{align}
    \mathcal B_{\mathrm{act}}(\ell)
    &\equiv
    \mathcal B(\ell,s_\ell)
    =
    \frac{\eta(\ell)}
         {\sqrt{2(\ell+2)}},
    \label{eq:comp_benchmark_B_active}
    \\
    \mathcal B_{\mathrm{dark}}(\ell)
    &\equiv
    \mathcal B(\ell,\chi_\perp)
    =
    1-\sqrt{\frac{2}{\ell+2}}.
    \label{eq:comp_benchmark_B_dark}
\end{align}
For the flat-band-dark state,
$\mathcal P_0(k)\ket{\chi_\perp}=0$, so its spectral weight is entirely
dispersive. These expressions are consistent with the weak-limit
analysis of continuously deformed three-state Grover walks~\cite{stefanak_limit_2014} 
and with the ballistic spreading of
one-dimensional lackadaisical quantum walks~\cite{wang_one-dimensional_2017}.

\begin{figure}[tbp]
    \centering
    \includegraphics[width=\linewidth]{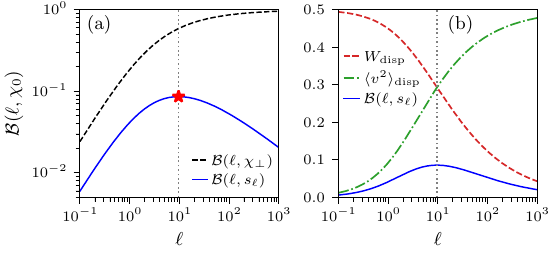}
    \caption{Ballistic coefficients of the restart-free LDTQW as a function of the
    self-loop weight $\ell$.
    (a) The blue solid line denotes $\mathcal B_{\mathrm{act}}(\ell)$ from 
    Eq.~\eqref{eq:comp_benchmark_B_active}, while the black dashed line represents 
    $\mathcal B_{\mathrm{dark}}(\ell)$ from Eq.~\eqref{eq:comp_benchmark_B_dark}.
    The red star marks the
    maximum of $\mathcal B_{\mathrm{act}}$ at
    $\ell_\ast=4(1+\sqrt{2})$.
    (b) Dispersive spectral weight
    $W_{\mathrm{disp}}(\ell,s_\ell)$ (red dashed line, Eq.~\eqref{eq:W_disp_sl}) 
    and conditional
    mean-square group velocity
    $\expval{v^2}_{\mathrm{disp}}$ (green dash-dotted, Eq.~\eqref{eq:active_v2_disp} ) 
    for the
    flat-band-active state.
    }
    \label{fig:ballistic_competition}
\end{figure}

The two coefficients have qualitatively different dependence on the
self-loop weight, as shown in
Fig.~\ref{fig:ballistic_competition}(a).
The flat-band-dark coefficient
$\mathcal B_{\mathrm{dark}}(\ell)$ increases monotonically with $\ell$.
In contrast, $\mathcal B_{\mathrm{act}}(\ell)$ is nonmonotonic: it
vanishes for both $\ell\to0^+$ and $\ell\to\infty$ and reaches a unique
maximum at
$\ell_\ast=4(1+\sqrt{2})$, where
$\mathcal B_{\mathrm{act}}(\ell_\ast)=3/2-\sqrt{2}$.

The origin of this nonmonotonicity can be understood by separating the
two ingredients entering the ballistic coefficient: the spectral weight
carried by the dispersive bands and their characteristic propagation speed.
For the flat-band-active state,
the former is
\begin{align}
W_{\mathrm{disp}}(\ell,s_\ell)
&=1-W_{\mathrm{fb}}(\ell,s_\ell)
=1-w_{\mathrm{act}}(\ell) \nonumber \\
&=\frac{1}{2}\big[ 1-\eta(\ell) \big],
\label{eq:W_disp_sl}
\end{align}
see Eqs.~\eqref{eq:lqw_flat_eigenvalues}--\eqref{eq:lqw_flat_weight_general}.
To isolate the latter, we normalize the $v^2(k)$-weighted dispersive
spectral contribution in Eq.~\eqref{eq:comp_ballistic_coeff} by the
total dispersive weight. This defines the conditional mean-square group
velocity
\begin{align}
\expval*{ v^2}_{\mathrm{disp}}
&\equiv
\frac{
\frac{1}{2\pi}\int_{-\pi}^{\pi}
dk\,v^2(k) \, w_{\mathrm{disp}}(k,s_\ell)
}{
W_{\mathrm{disp}}(\ell,s_\ell)
}
=
\frac{\mathcal B(\ell,s_\ell)}
{W_{\mathrm{disp}}(\ell,s_\ell)}
\nonumber\\
&=
\frac{1}{2}
\left[
1-\sqrt{\frac{2}{\ell+2}}
\right] ,
\label{eq:active_v2_disp}
\end{align}
which is the mean value of $v^2(k)$ over the dispersive part of the
initial state's spectral decomposition, conditioned on the dispersive spectral sector. 
Consequently, the active-state ballistic coefficient factorizes exactly as
$\mathcal B_{\mathrm{act}}
=
W_{\mathrm{disp}}\expval{v^2}_{\mathrm{disp}}$.

As shown in Fig.~\ref{fig:ballistic_competition}(b),
$\expval{v^2}_{\mathrm{disp}}$ increases monotonically with $\ell$,
consistent with the increasing dispersive velocities discussed in
Sec.~\ref{subsec:LDTQW_spectrum}, whereas
$W_{\mathrm{disp}}(\ell,s_\ell)$ decreases monotonically to zero.
The maximum of $\mathcal B_{\mathrm{act}}(\ell)$ therefore arises from
an exact competition between increasingly rapid propagation within the
dispersive sector and the simultaneous depletion of its spectral
weight.

\begin{figure}[tb]
    \centering
    \includegraphics[width=\linewidth]{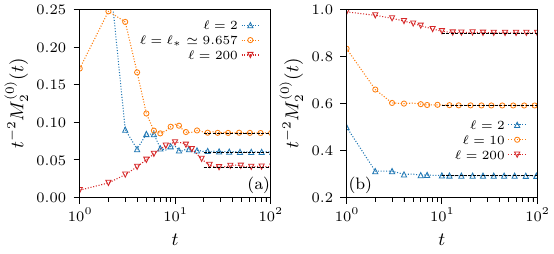}
    \caption{Scaled MSD of the restart-free LDTQW.
    (a) Flat-band-active state. The black dashed line denotes 
    $\mathcal B_{\mathrm{act}}(\ell)$ from Eq.~\eqref{eq:comp_benchmark_B_active}. 
    (b) Flat-band-dark state. The black dashed line denotes 
    $\mathcal B_{\mathrm{dark}}(\ell)$ from Eq.~\eqref{eq:comp_benchmark_B_dark}. 
    For both panels, symbols are obtained from numerical simulation and the dotted 
    lines are guides to the eye.}
    \label{fig:restart_free_msd}
\end{figure}

Figure~\ref{fig:restart_free_msd} illustrates the approach to the
ballistic regime by plotting the scaled MSD $t^{-2}M_2^{(0)}(t)$.
For both benchmark states, the scaled MSD approaches the corresponding
asymptotic coefficient on a time scale of order $10$ steps, with the
remaining finite-time corrections becoming small within a few tens of
steps. Panel~\ref{fig:restart_free_msd}(a) also makes the nonmonotonic
dependence of $\mathcal B_{\mathrm{act}}(\ell)$ on $\ell$ directly visible:
the asymptotic value at $\ell=\ell_\ast$ exceeds those at both smaller
and much larger self-loop weights. In contrast, panel
\ref{fig:restart_free_msd}(b) reflects the monotonic increase of
$\mathcal B_{\mathrm{dark}}(\ell)$ with $\ell$.

\section{Stochastic restart of the LDTQW}
\label{sec:single_site_restart}

We now interrupt the restart-free unitary dynamics at random times and
reinitialize the complete walker-coin state. Stochastic restart and its
renewal formulation have been studied extensively in classical and
quantum settings~\cite{evans_stochastic_2020,wald_classical_2021,
wald_stochastic_2025,chelminiak_discrete-time_2025}.
Here we formulate the dynamics independently of the inter-restart
distribution and subsequently specialize the resulting renewal
relations to geometric and power-law restart in
Secs.~\ref{sec:geometric_restart} and~\ref{sec:powerlaw_restart}.

We consider single-site restart to the fixed state
$\ket{\Psi_R}\equiv\ket{\chi_R}\otimes\ket{n_R}$ and take it to be identical 
to the initial state, i.e., 
$\ket{\Psi(0)}=\ket{\Psi_R}$. Each renewal interval therefore starts
from the same walker-coin configuration and evolves according to the
restart-free LDTQW of Sec.~\ref{sec:LDTQW}.

\subsection{Restart protocol and renewal statistics}
\label{subsec:single_restart_protocol}

Let $\mathcal T_1$, $\mathcal T_2$, $\ldots$ denote the waiting-time 
intervals between successive restarts. We assume that they are independent
and identically distributed positive-integer random variables,
independent of the quantum-walk dynamics, with probability 
$p_m\equiv\Pr(\mathcal T_j=m)$ for $m\geq1$ and $p_0=0$.
Starting from $\ket{\Psi_R}$, the walker evolves unitarily for
${\mathcal T}_1$ steps and is then instantaneously reinitialized to the same state.
It subsequently evolves for ${\mathcal T}_2$ steps 
before the next restart, and so on. 
The $j$th restart occurs at
$t_j^{\mathrm R}=\sum_{\nu=1}^{j}\mathcal T_\nu$.
A restart at $t_j^{\mathrm R}$ is applied immediately after the
corresponding unitary step, so the age immediately after a restart is
zero.

The waiting-time survival probability is defined as
$\Phi_a\equiv\Pr({\mathcal T}>a)
=\sum_{m=a+1}^{\infty}p_{m}$, with
$a=0,1,2,\ldots$ and $\Phi_0=1$. Thus, $\Phi_a$ is the
probability that an inter-restart interval survives for at least $a$
complete time steps without another restart.
We denote by $u_j$, $j\geq1$, the renewal mass, i.e., the
probability that a restart occurs exactly at time $j$, irrespective of
earlier restarts. For convenience, we set $u_0=1$,
which identifies $t=0$ as the renewal origin but does not represent a
physical restart. The renewal mass then satisfies
$u_j=\sum_{\nu=0}^{j}u_\nu p_{j-\nu}$ for $j\geq1$.

For any sequence $f_n$, let 
$\widetilde f(z) \equiv \sum_{n=0}^{\infty}f_n z^n$ denote its generating function. 
The generating functions of $\Phi_a$ and $u_j$ may then be
written in terms of the generating function of $p_m$ as
\begin{align}
    \widetilde\Phi(z)
    &=
    \frac{1-\widetilde p(z)}{1-z}, \quad \text{and} \quad 
    \widetilde u(z)
    =
    \frac{1}{1-\widetilde p(z)} ,
    \label{eq:renewal_generating}
\end{align}
respectively.

The renewal variable relevant to the restart dynamics is the age
$A_t$, defined as the number of steps elapsed since the most recent
restart at observation time $t$, or since the initial preparation if no
restart has yet occurred. Its exact finite-time distribution is
\begin{align}
\mathcal A_t(a)
&\equiv
\Pr(A_t=a)
=
u_{t-a}\Phi_a,
\qquad
0\leq a\leq t.
\label{eq:restart_age_distribution}
\end{align}
For $a<t$, this corresponds to a renewal at $t-a$ followed by $a$
restart-free steps; the boundary value $a=t$ describes the no-restart
history, with probability $\Phi_t$. The age distribution is normalized, i.e.,
$\sum_{a=0}^{t}\mathcal A_t(a)=1$.

\subsection{Renewal representation of observables}
\label{subsec:transient_restart}

Conditioned on age $A_t=a$, the walker has undergone $a$ uninterrupted
restart-free steps from $\ket{\Psi_R}$. 
The occupation probability conditioned on this age is therefore 
the restart-free quantity 
$P^{(0)}(n,a|n_R,\chi_R)$, given in
Eq.~\eqref{eq:bare_probability}. Averaging over the age distribution 
at observation time $t$ gives the
exact time-dependent site-occupation probability
$P(n,t|n_R,\chi_R)$ under stochastic restart:
\begin{align}
    P(n,t|n_R,\chi_R)
    &=
    \sum_{a=0}^{t}
    \mathcal A_t(a)
    P^{(0)}(n,a|n_R,\chi_R)
    \nonumber \\
    &=
    \sum_{a=0}^{t}
    u_{t-a}\Phi_a
    P^{(0)}(n,a|n_R,\chi_R).
    \label{eq:single_restart_renewal}
\end{align}
Thus, all dependence on the restart protocol enters through the age
weights $\mathcal A_t(a)$, while the dynamics within a renewal interval
is completely determined by the restart-free walk.

The same renewal representation applies to 
any time-dependent restart-free observable $O^{(0)}(t)$ 
whose value after a restart depends only on the 
elapsed restart-free evolution time. 
The corresponding restart-averaged quantity reads 
$O(t)=\sum_{a=0}^{t}\mathcal A_t(a) O^{(0)}(a)$.
Equations~\eqref{eq:renewal_generating}--\eqref{eq:restart_age_distribution} give
\begin{align}
    \widetilde O(z)
    =
    \frac{1}{1-\widetilde p(z)}
        \sum_{a=0}^{\infty} z^a 
        \Phi_a O^{(0)}(a) ,
    \label{eq:general_renewal_transform}
\end{align}
which provides a general generating-function representation of
the restart dynamics.

In the long-time limit, provided that the mean waiting time
$\langle{\mathcal T}\rangle$ is finite and that the integer-valued 
waiting-time distribution $p_m$ is aperiodic,  or equivalently has span one, 
the discrete renewal theorem gives
$u_j\to1/\langle{\mathcal T}\rangle$ as $j\to\infty$~\cite{erdos_property_1949, feller_introduction_2009}. 
The waiting-time
laws (geometric and power-law) considered in the following have support
on every positive integer and therefore satisfy this condition.
Equation~\eqref{eq:restart_age_distribution}
then yields the stationary age distribution
\begin{align}
\pi_a\equiv\lim_{t\to\infty}\mathcal A_t(a)
=\frac{\Phi_a}{\langle{\mathcal T}\rangle}.
\label{eq:pi-def-gen}
\end{align}
The tail-sum identity
$\sum_{a=0}^{\infty}\Phi_a=\langle{\mathcal T}\rangle$ ensures that
$\pi_a$ is normalized, i.e., 
$\sum_{a=0}^{\infty}\pi_a=1$.

Whenever this stationary-age distribution exists, the stationary 
site-occupation probability of the walker under restart 
is obtained from Eq.~\eqref{eq:single_restart_renewal} as 
\begin{align}
    P^{\mathrm{st}}(n|n_R,\chi_R)
    &\equiv
    \lim_{t\to\infty}P(n,t|n_R,\chi_R)
    \nonumber \\
    &=
    \frac{1}{\langle{\mathcal T}\rangle}
    \sum_{a=0}^{\infty}
    \Phi_a
    P^{(0)}(n,a|n_R,\chi_R) .
    \label{eq:general_restart_stationary}
\end{align}
For more general observables, the corresponding stationary renewal
average requires convergence of the age-weighted sum. In particular,
the existence of a normalized stationary site-occupation distribution
does not imply that all of its spatial moments are finite.

For the mean-squared displacement from the restart site, conditioning
on the age gives
\begin{align}
M_2(t)
&=
\sum_{a=0}^{t}
\mathcal A_t(a)
M_2^{(0)}(a),
\label{eq:comp_moment_renewal}
\end{align}
where $M_2^{(0)}(a)$ is the restart-free MSD of
Eq.~\eqref{eq:comp_second_moments_r0} with
$n_0=n_R$ and $\chi_0=\chi_R$.
When a stationary age distribution
exists, a finite stationary MSD further requires convergence of the
corresponding age-weighted sum of
$M_2^{(0)}(a)$. 
This distinction will become central for power-law
restart, for which the conditions for a stationary occupation profile
and for a finite stationary MSD are different.

Equations~\eqref{eq:single_restart_renewal},~\eqref{eq:general_renewal_transform},~\eqref{eq:general_restart_stationary}, and~\eqref{eq:comp_moment_renewal} provide the protocol-independent
framework used throughout the remainder of the analysis. In the
following sections we apply them to geometric and power-law restart.

\section{LDTQW with geometric restart}
\label{sec:geometric_restart}

We first apply the renewal framework of
Sec.~\ref{sec:single_site_restart} to geometric restart. In this case
the restart probability is constant at each time step, introducing a
single characteristic renewal time scale. The flat-band-active and flat-band-dark
coin states introduced in Sec.~\ref{subsec:LDTQW_benchmark_states} will
serve as contrasting restart states.

The inter-restart times are geometrically distributed,
\begin{align}
    p_m^{(\mathrm g)}
    &=
    q(1-q)^{m-1},
    \qquad
    0<q<1,
    \label{eq:waiting_geometric}
\end{align}
where $q$ is the restart probability per time step. The mean waiting time
is $\langle{\mathcal T}\rangle_{\mathrm g}=1/q$, while the survival probability
is $\Phi_a^{(\mathrm g)} =(1-q)^a$. Owing to the memoryless character of
the geometric distribution, the renewal mass is
$u_j^{(\mathrm g)}=q$ for every $j\geq1$.

\subsection{Site-occupation probability}
\label{subsec:geometric_occup}

Substitution into the general renewal relation
\eqref{eq:single_restart_renewal} gives the exact finite-time
site-occupation probability
\begin{align}
    P_{\mathrm g}(n,t|n_R,\chi_R)
    &=
    (1-q)^t
    P^{(0)}(n,t|n_R,\chi_R)
    \nonumber\\
    &\quad+
    q\sum_{a=0}^{t-1}
    (1-q)^a
    P^{(0)}(n,a|n_R,\chi_R).
    \label{eq:geometric_transient}
\end{align}
The first term represents the no-restart history, while the remaining
terms collect histories according to the age of the most recent
restart. Thus Eq.~\eqref{eq:geometric_transient} describes the
transient dynamics as an exponentially weighted sampling of
restart-free walks of different ages. At fixed $t$, it continuously
recovers the restart-free occupation probability as $q\to0$.

Since the mean $\langle{\mathcal T}\rangle_{\mathrm g}$ is finite for every $q>0$,
the renewal age approaches the normalized stationary distribution
$\pi_a^{(\mathrm g)} =q(1-q)^a$, see Eq.~\eqref{eq:pi-def-gen}. 
Equation~\eqref{eq:general_restart_stationary}
then gives the exact stationary site-occupation profile
\begin{align}
    P_{\mathrm g}^{\mathrm{st}}(n|n_R,\chi_R)
    &=
    q\sum_{a=0}^{\infty}
    (1-q)^a
    P^{(0)}(n,a|n_R,\chi_R).
    \label{eq:geometric_stationary}
\end{align}
Increasing $q$
suppresses long restart-free intervals more strongly and consequently
confines the stationary distribution more tightly around the restart
site.

The distinction between intrinsic flat-band localization and
restart-induced confinement is especially transparent at the restart site.
For the flat-band-active state $\ket{s_\ell}$, the restart-free return
probability approaches the nonzero value $w_{\mathrm{act}}^2(\ell)$ according to
Eq.~\eqref{eq:lqw_active_return}. As $q\to0$, the stationary renewal
average samples progressively longer restart-free intervals. 
Equation~\eqref{eq:geometric_stationary} therefore gives
\begin{align}
    \lim_{q\to0}
    P_{\mathrm g}^{\mathrm{st}}(n_R|n_R,s_\ell)
    &=
    w_{\mathrm{act}}^2(\ell).
    \label{eq:geom_active_origin_limit}
\end{align}
Hence the occupation of the restart site remains finite even in the
weak-restart limit, reflecting the intrinsic localization already present
in the restart-free active-state dynamics.

The flat-band-dark state $\ket{\chi_\perp}$ behaves qualitatively
differently. Its flat-band projection vanishes, and its restart-free
return probability decays algebraically according to
Eq.~\eqref{eq:lqw_dark_return}. Combining this asymptotic behavior with
Eq.~\eqref{eq:geometric_stationary} gives
\begin{align}
    P_{\mathrm g}^{\mathrm{st}}(n_R|n_R,\chi_\perp)
    &\sim
    \frac{1}{\pi}
    \sqrt{\frac{2}{\ell}}\,
    q\ln\frac{1}{q},
    \qquad
    q\to0 .
    \label{eq:geom_dark_origin_asymptotic}
\end{align}
Hence the stationary restart-site occupation vanishes in the
weak-restart limit. Since the flat-band component is identically absent,
the finite-$q$ stationary confinement of this state is entirely
restart-induced.
The logarithmic factor in Eq.~\eqref{eq:geom_dark_origin_asymptotic} 
follows directly from the algebraic
$a^{-1}$ restart-free return. Under geometric restart, this tail enters
the stationary average through
$q\sum_{a=1}^{\infty}(1-q)^a/a=q\ln(1/q)$. Thus the weak-restart asymptote
retains a direct
signature of the slow dispersive return of the restart-free walk.

\begin{figure}[tbp]
    \centering
    \includegraphics[width=1\linewidth]{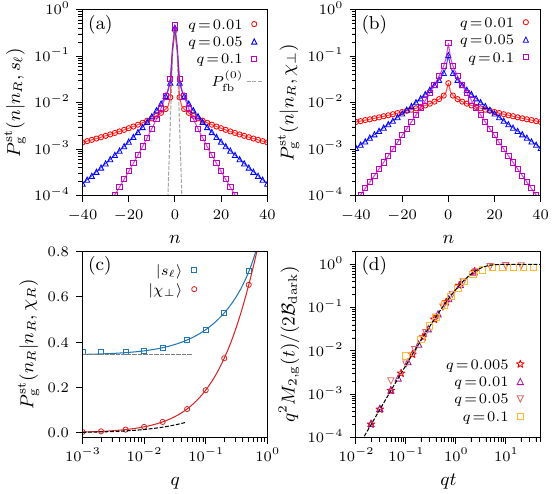}
    \caption{
    Geometric restart for $\ell=2$ and $n_R=0$.
    (a) and (b) Stationary site-occupation profiles for the flat-band-active
    state $\ket{s_\ell}$ and the flat-band-dark state $\ket{\chi_\perp}$, 
    respectively. Solid curves are stationary renewal
    results from Eq.~\eqref{eq:geometric_stationary} and symbols are 
    numerical results averaged over $5\times 10^5$ stochastic realizations. 
    The gray dashed 
    curve in (a) is the intrinsic flat-band profile
    $P_{\mathrm{fb}}^{(0)}(n|n_R,s_\ell)$ from
    Eq.~\eqref{eq:lqw_active_localized_profile}.
    (c) Stationary restart-site occupation probability. Solid lines are 
    obtained from Eq.~\eqref{eq:geometric_stationary}, while symbols 
    denote numerical results averaged over $5\times 10^5$ stochastic 
    realizations. Dashed lines in gray and black show
    the weak-restart limits~\eqref{eq:geom_active_origin_limit} and
    \eqref{eq:geom_dark_origin_asymptotic}, respectively.
    (d) Weak-restart MSD scaling for the flat-band-dark state. Symbols show
    $q^2M_{2,\rm g}(t)/(2 \mathcal B_{\mathrm{dark}})$ versus $qt$ from 
    numerical stochastic realizations, while the dashed black 
    curve  is $1-(1+qt)e^{-qt}$ from
    Eq.~\eqref{eq:geom_msd_crossover}.
    }
    \label{fig:geometric_restart_competition}
\end{figure}

Figures~\ref{fig:geometric_restart_competition}(a) and
\ref{fig:geometric_restart_competition}(b) show the corresponding
stationary profiles for the flat-band-active and flat-band-dark states, 
respectively. For both restart states,
increasing $q$ suppresses long excursions and narrows the distribution
around the restart site. The two states nevertheless retain markedly
different spatial structures. For $\ket{s_\ell}$, the intrinsic
flat-band contribution $P_{\mathrm{fb}}^{(0)}(n|n_R,s_\ell)$ produces 
a pronounced localized core, shown by
the dashed gray curve in panel (a) as a restart-free reference. For
$\ket{\chi_\perp}$ this intrinsic contribution is absent, so the
stationary confinement is generated entirely by restart of the
otherwise dispersive walk.
The comparison also shows that intrinsic localization does not enhance
the active-state profile uniformly in space. The flat-band-active 
state has the
larger occupation near the restart site, whereas farther from the
localized core the dark-state profile becomes larger. This behavior is
consistent with the different spectral content of the two benchmark
states: a finite fraction of the active-state weight is concentrated in
the flat band, while the flat-band-dark state is entirely dispersive 
and therefore
develops broader restart-confined tails.

Panel~\ref{fig:geometric_restart_competition}(c) isolates the same
contrast locally at the restart site. As $q\to0$, the active-state 
occupation approaches the
finite intrinsic value $w_{\mathrm{act}}^2(\ell)$, whereas the 
flat-band-dark state occupation
vanishes according to the $q\ln(1/q)$ law in
Eq.~\eqref{eq:geom_dark_origin_asymptotic}. The visible deviation of the
dark-state data from the leading asymptotic form over the range shown is
consistent with its logarithmically slow convergence.

\subsection{Mean-squared displacement}
\label{subsec:geometric_msd}

We next turn from the stationary spatial profile to the global spread of
the walker. In the absence of restart, the MSD grows
ballistically according to Eq.~\eqref{eq:comp_bare_ballistic_r0}.
Geometric restart interrupts these ballistic excursions on the
characteristic time scale $q^{-1}$.

Substituting the geometric age probabilities into the general MSD
renewal relation~\eqref{eq:comp_moment_renewal} gives the exact transient
MSD
\begin{align}
    M_{2,\mathrm g}(t)
    &=
    (1-q)^tM_2^{(0)}(t)
    +
    q\sum_{a=0}^{t-1}
    (1-q)^aM_2^{(0)}(a).
    \label{eq:geom_msd_transient}
\end{align}
The first term corresponds to the
no-restart history, while the sum averages over trajectories whose most
recent restart occurred at or before time $t$. At fixed observation time,
Eq.~\eqref{eq:geom_msd_transient} reduces continuously to
$M_2^{(0)}(t)$ as $q\to0$.

The interplay between ballistic propagation and restart becomes
particularly transparent in the weak-restart scaling regime. Using the
restart-free asymptotic form~\eqref{eq:comp_bare_ballistic_r0} in
Eq.~\eqref{eq:geom_msd_transient}, and taking $q\to0$ and
$t\to\infty$ with $qt$ fixed, gives (see Appendix~\ref{app:geometric_msd})
\begin{align}
    M_{2,\mathrm g}(t)
    &\sim
    \frac{2\mathcal B(\ell,\chi_R)}{q^2}
    \left[
        1-(1+qt)e^{-qt}
    \right].
    \label{eq:geom_msd_crossover}
\end{align}
For $qt\ll1$, the scaling function behaves as
$1-(1+qt)e^{-qt}\sim(qt)^2/2$, recovering the restart-free ballistic
law $M_{2,\mathrm g}(t)\sim\mathcal B(\ell,\chi_R)t^2$. For
$qt\gg1$, the scaling function approaches unity. The crossover
therefore occurs on the renewal time scale $t\sim q^{-1}$ and separates
an initial restart-free ballistic regime from long-time confinement.

For every fixed $q>0$, the exponential stationary-age distribution also
ensures a finite stationary MSD,
$M_{2,\mathrm g}^{\mathrm{st}}
=q\sum_{a=0}^{\infty}(1-q)^aM_2^{(0)}(a)$.
Using the restart-free ballistic scaling~\eqref{eq:comp_bare_ballistic_r0},
we obtain
\begin{align}
    M_{2,\mathrm g}^{\mathrm{st}}
    &\approx
    \mathcal B(\ell,\chi_R)
    \frac{(1-q)(2-q)}{q^2}.
    \label{eq:geom_stationary_msd_asymptotic}
\end{align}
The factor $(1-q)(2-q)/q^2=\langle a^2\rangle_{\mathrm g}$ is the
second moment of the stationary geometric age distribution. Hence, in
the weak-restart limit,
$M_{2,\mathrm g}^{\mathrm{st}}\sim
2\mathcal B(\ell,\chi_R)/q^2$, and the corresponding root-mean-square
width scales as $q^{-1}$. Physically, geometric restart limits the
duration of uninterrupted ballistic spreading to a characteristic age
of order $q^{-1}$, thereby converting ballistic growth into a finite
stationary spatial scale. This inverse-restart-probability scaling is
the discrete-time analogue of the inverse-restart-rate length scale in
finite-speed telegraphic motion~\cite{masoliver_telegraphic_2019}.

Figure~\ref{fig:geometric_restart_competition}(d) verifies this
crossover for the flat-band-dark state. Measuring time in units of
$q^{-1}$ and the MSD in units of
$2\mathcal B_{\mathrm{dark}}/q^2$ collapses the data onto the scaling
function $1-(1+qt)e^{-qt}$, connecting the ballistic small-$qt$
regime to the weak-restart stationary large-$qt$ limit. The collapse
therefore makes explicit the single renewal time scale governing the
transition from restart-free propagation to restart-induced confinement.

Geometric restart thus provides a particularly simple reference
case: the transient dynamics crosses over on the finite time scale
$q^{-1}$ to a normalized stationary spatial profile with a finite MSD
for every $q>0$. The local structure of this stationary profile retains a
clear signature of whether the restart coin state populates the intrinsic
flat-band sector. As we show next, this picture changes qualitatively
for power-law restart, where algebraically distributed renewal
intervals introduce distinct thresholds for stationarity and moment
convergence.

\section{LDTQW with power-law restart}
\label{sec:powerlaw_restart}

We now turn to power-law restart, for which the inter-restart waiting times
are broadly distributed with no intrinsic cutoff scale. 
This leads to a qualitatively richer long-time behavior than
for geometric restart. In particular, the threshold for a normalized stationary
site-occupation distribution differs from the thresholds for convergence
of its spatial moments.

We consider the power-law waiting-time distribution
\begin{align}
    p_m
    &=
    \frac{m^{-s}}{\zeta(s)},
    \qquad
    s>1,
    \label{eq:waiting_powerlaw}
\end{align}
where $\zeta(s)$ is the Riemann zeta function. The mean waiting time is
$\langle{\mathcal T}\rangle_{\mathrm{pl}}
=\zeta(s-1)/\zeta(s)$ for $s>2$, whereas it diverges for
$1<s\leq2$. 

The corresponding survival probability is
\begin{align}
    \Phi_a^{(\mathrm{pl})}
    &=
    \frac{\zeta(s,a+1)}{\zeta(s)}
    \sim
    \frac{a^{1-s}}
         {(s-1)\zeta(s)},
    \qquad
    a\rightarrow\infty,
    \label{eq:powerlaw_survival}
\end{align}
where $\zeta(s,a)$ denotes the Hurwitz zeta function. In contrast to
the exponential age suppression produced by geometric restart,
Eq.~\eqref{eq:powerlaw_survival} assigns an algebraically decaying
weight to long restart-free intervals.
The waiting-time generating function is
$\widetilde p_{\mathrm{pl}}(z)
=\operatorname{Li}_s(z)/\zeta(s)$, where
$\operatorname{Li}_s(z)$ is the polylogarithm. From
Eq.~\eqref{eq:renewal_generating}, the corresponding renewal mass is
determined by its generating function 
$\widetilde u_{\mathrm{pl}}(z)
=\zeta(s)/[\zeta(s)-\operatorname{Li}_s(z)]$.

\subsection{Site-occupation probability}
\label{subsec:powerlaw_occup}

Substituting the survival probability into the general renewal
relation~\eqref{eq:single_restart_renewal} gives the exact finite-time
site-occupation probability
\begin{align}
    P_{\rm pl}(n,t|n_R,\chi_R)
    &=
    \frac{1}{\zeta(s)}
    \sum_{a=0}^{t}
    u_{t-a}^{(\mathrm{pl})}
    \zeta(s,a+1)
    \nonumber\\
    &\hskip50pt \times
    P^{(0)}(n,a|n_R,\chi_R).
    \label{eq:powerlaw_transient}
\end{align}
Equation~\eqref{eq:powerlaw_transient} is valid for the entire range
$s>1$, including the infinite-mean regime $1<s\leq2$. The distinction
between finite- and infinite-mean restart therefore emerges only in the
long-time behavior.

The mean  $\langle{\mathcal T}\rangle_{\mathrm{pl}}$ is finite only for $s>2$.
Hence, the renewal age approaches the normalized stationary 
distribution~\eqref{eq:pi-def-gen} only in this regime:
\begin{align}
    \pi_a^{(\mathrm{pl})}
    &=
    \frac{\zeta(s,a+1)}{\zeta(s-1)}
    \sim
    \frac{a^{1-s}}
         {(s-1)\zeta(s-1)},
    \qquad
    a\rightarrow\infty .
    \label{eq:powerlaw_stationary_age}
\end{align}
For $s>2$, Eq.~\eqref{eq:general_restart_stationary} therefore gives
the stationary site-occupation distribution 
\begin{align}
    P_{\rm pl}^{\mathrm{st}}(n|n_R,\chi_R)
    &=
    \sum_{a=0}^{\infty}
    \frac{\zeta(s,a+1)}{\zeta(s-1)}
    P^{(0)}(n,a|n_R,\chi_R) .
    \label{eq:powerlaw_stationary}
\end{align}
Increasing $s$ suppresses long restart-free intervals more
strongly and consequently produces tighter confinement around the restart
site.

\begin{figure}[tbp]
    \centering
    \includegraphics[width=1\linewidth]{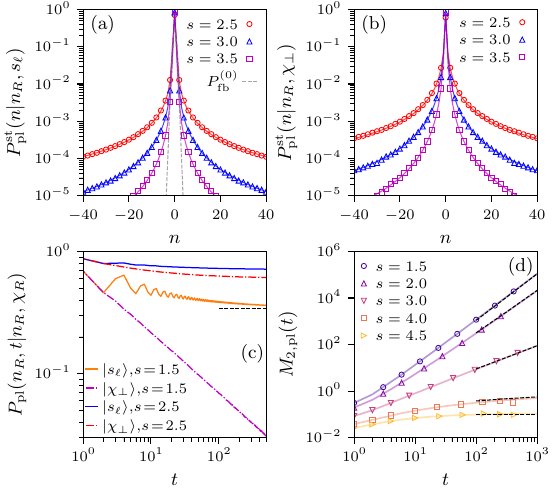}
    \caption{
    Power-law restart for $\ell=2$ and $n_R=0$.
    (a) and (b) Stationary site-occupation profiles for the flat-band-active
    state $\ket{s_\ell}$ and the flat-band-dark state
    $\ket{\chi_\perp}$, respectively. 
    Solid lines are stationary renewal results from 
    Eq.~\eqref{eq:powerlaw_stationary} and symbols are numerical 
    results averaged over $5\times 10^5$ stochastic realizations.
    The dashed gray curve in (a) is the
    intrinsic flat-band contribution $P_{\mathrm{fb}}^{(0)}(n|n_R,s_\ell)$ from
    Eq.~\eqref{eq:lqw_active_localized_profile}.
    (c) Restart-site occupation probability as a function of time. 
    The horizontal black dashed line denotes $w_{\rm act}^2(\ell=2)\approx0.343$; 
    the remaining curves are numerical results averaged 
    over $5\times10^5$ stochastic realizations. 
    (d) MSD for the flat-band-dark state. Symbols show results 
    from numerical stochastic
    simulations, while colored solid lines are the exact renewal 
    results from Eq.~\eqref{eq:powerlaw_msd_transient}. 
    For $s\leq4$, the black dashed lines denote the long-time asymptotic
    forms in Eq.~\eqref{eq:comp_powerlaw_scaling}, while for $s>4$ the
    black dashed line denotes the ballistic approximation to the stationary
    MSD in Eq.~\eqref{eq:comp_powerlaw_scaling_msd_steady}.    }
    \label{fig:powerlaw_restart}
\end{figure}

Figures~\ref{fig:powerlaw_restart}(a) and
\ref{fig:powerlaw_restart}(b) show the stationary profiles for the
flat-band-active and flat-band-dark states, respectively. Increasing
$s$ reduces the contribution of long renewal ages and narrows both
distributions around the restart site, while their spatial structures
remain distinct. For $\ket{s_\ell}$, the flat-band contribution 
$P_{\mathrm{fb}}^{(0)}(n|n_R,s_\ell)$ produces an
intrinsic localized core, shown in panel (a) by the dashed gray 
curve, whereas this contribution is absent for
$\ket{\chi_\perp}$. Consequently, the active state is more strongly
concentrated near the restart site, while the fully dispersive
flat-band-dark state develops broader restart-averaged tails.

The situation changes qualitatively for $1<s\leq2$. In this regime
the mean waiting time diverges and no normalized stationary age
distribution exists. As shown in Appendix~\ref{app:powerlaw_msd}, the
renewal age satisfies $A_t\to\infty$ in probability as $t\to\infty$.
Since, at every fixed lattice site, the restart-free occupation
approaches its flat-band contribution at large age, the renewal average
inherits the same fixed-site limit:
\begin{align}
    P_{\rm pl}(n,t|n_R,\chi_R)
    &\longrightarrow
    P_{\mathrm{fb}}^{(0)}(n|n_R,\chi_R),
    \quad
    1<s\leq2 .
    \label{eq:comp_powerlaw_pointwise}
\end{align}
This convergence is pointwise and does not define a normalized
stationary site-occupation distribution. 
For the flat-band-active state, the limiting profile is the nonzero
intrinsic flat-band profile and carries only the flat-band weight
$w_{\mathrm{act}}(\ell)<1$.
The complementary probability is
carried to increasingly large distances. For the flat-band-dark state,
$P_{\mathrm{fb}}^{(0)}(n|n_R,\chi_\perp) = 0$, so the occupation of every
fixed lattice site vanishes as $t\to\infty$.
Figure~\ref{fig:powerlaw_restart}(c) depicts the limiting behavior
\eqref{eq:comp_powerlaw_pointwise} at the restart site $n_R$.
For $s=1.5$, as time progresses, the restart-site occupation probability
for the flat-band-active state approaches its flat-band limit
$w_{\rm act}^2$ given in Eq.~\eqref{eq:lqw_active_return}, while that
for the flat-band-dark state decays to zero. In contrast, for $s=2.5$,
the restart-site occupation probabilities for both states approach
nonzero stationary values determined by
Eq.~\eqref{eq:powerlaw_stationary}.
Physically, renewal aging makes the walk increasingly likely to be
observed during a very old restart-free interval. The dispersive
component has then propagated beyond any fixed spatial window,
whereas the flat-band component remains localized, allowing local
spectral memory to survive without global stationarity.

The exponent $s=2$ thus marks the threshold for the existence of a
normalized stationary site-occupation distribution. Importantly, this
condition concerns normalization of the long-time spatial profile and
does not by itself guarantee convergence of its spatial moments. We 
turn next to the second moment, whose convergence introduces the
distinct finite-MSD threshold at $s=4$.

\subsection{Mean-squared displacement}
\label{subsec:powerlaw_msd}

The exact transient MSD follows directly from the general renewal
relation~\eqref{eq:comp_moment_renewal}. For power-law restart, we thus have 
\begin{align}
    M_{2,\rm pl}(t)
    &=
    \frac{1}{\zeta(s)}
    \sum_{a=0}^{t}
    u_{t-a}^{(\mathrm{pl})}
    \zeta(s,a+1)
    M_2^{(0)}(a).
    \label{eq:powerlaw_msd_transient}
\end{align}
Thus the MSD samples the restart-free spreading over the same broad
distribution of renewal ages that governs the occupation probability.
Because the restart-free walk is ballistic,
$M_2^{(0)}(a)\sim\mathcal B(\ell,\chi_R)a^2$, the heavy tail of the
renewal statistics produces several distinct long-time regimes.

Using the restart-free ballistic asymptotics in
Eq.~\eqref{eq:powerlaw_msd_transient}, one obtains (see Appendix~\ref{app:powerlaw_msd})
\begin{align}
    M_{2,\rm pl}(t)
    &\sim
    \begin{cases}
    \dfrac{\mathcal B(\ell,\chi_R)}{2}
    (2-s)(3-s)t^2,
    &1<s<2,
    \\[3mm]
    \dfrac{\mathcal B(\ell,\chi_R)}{2}
    \dfrac{t^2}{\ln t},
    &s=2,
    \\[3mm]
    \dfrac{\mathcal B(\ell,\chi_R)}
    {(s-1)(4-s)\zeta(s-1)}
    t^{4-s},
    &2<s<4,
    \\[3mm]
    \dfrac{\mathcal B(\ell,\chi_R)}
    {3\zeta(3)}
    \ln t,
    &s=4 .
    \end{cases}
    \label{eq:comp_powerlaw_scaling}
\end{align}
For $1<s<2$, restart modifies the ballistic prefactor but not the
$t^2$ growth exponent. At $s=2$, the ballistic law acquires the
marginal logarithmic correction $t^2/\ln t$. For $2<s<4$, a
normalized stationary site-occupation distribution already exists,
yet the MSD continues to grow algebraically as $t^{4-s}$. At $s=4$,
this algebraic growth is replaced by the marginal logarithmic law.

The condition for convergence of the stationary MSD follows directly
from the large-age tail of the stationary age distribution. 
For $s>2$, using Eq.~\eqref{eq:powerlaw_stationary_age}, one may write 
\begin{align}
    \pi_a^{(\mathrm{pl})} M_2^{(0)}(a)
    &\sim
    \frac{\mathcal B(\ell,\chi_R)}
    {(s-1)\zeta(s-1)}
    a^{3-s}.
    \label{eq:powerlaw_stationary_msd_tail}
\end{align}
The stationary MSD therefore converges only for $s>4$. Thus $s=4$
marks the finite-MSD threshold, distinct from the stationarity
threshold at $s=2$: for $2<s\leq4$, the site-occupation distribution
is stationary, while its second moment remains divergent.

For $s>4$, the stationary-age average of the restart-free MSD is finite,
and the stationary second moment is
\begin{align}
    M_{2,\rm pl}^{\mathrm{st}}
    &=
    \frac{1}{\zeta(s-1)}
    \sum_{a=0}^{\infty}
    \zeta(s,a+1)M_2^{(0)}(a)
    \label{eq:slow-converge} \\
    &\approx
    \frac{
        \mathcal B(\ell,\chi_R)
        \left[
            2\zeta(s-3)
            -3\zeta(s-2)
            +\zeta(s-1)
        \right]
    }
    {6\zeta(s-1)} .
    \label{eq:comp_powerlaw_scaling_msd_steady}
\end{align}
The first line is the exact stationary renewal average, whereas the
second is obtained by replacing the restart-free MSD by its leading
ballistic form
$M_2^{(0)}(a)\sim\mathcal B(\ell,\chi_R)a^2$
throughout the renewal sum.

The resulting MSD hierarchy is illustrated in
Fig.~\ref{fig:powerlaw_restart}(d) for the flat-band-dark benchmark
state. The values $s=1.5$, $2$, $3$, $4$, and $4.5$ sample the
different long-time regimes. For $s=1.5$, the MSD remains ballistic;
at $s=2$, it follows the marginal $t^2/\ln t$ law. For $s=3$, the
site-occupation probability has already become stationary, while the
MSD continues to grow linearly. At $s=4$, the growth is logarithmic,
whereas for $s=4.5$ the MSD approaches a finite stationary value.
The flat-band-active and flat-band-dark states have the same asymptotic exponents and
thresholds; their leading amplitudes differ through the ballistic
coefficient $\mathcal B(\ell,\chi_R)$. 

The stationary MSD becomes singular as the second-moment threshold is
approached from above. The large-age ballistic contribution gives
\begin{align}
    M_{2,\rm pl}^{\mathrm{st}}
    &\sim
    \frac{\mathcal B(\ell,\chi_R)}
    {3\zeta(3)}
    \frac{1}{s-4},
    \qquad
    s\to4^+ ,
    \label{eq:comp_powerlaw_stationary_divergence}
\end{align}
which implies that the stationary second moment diverges as $(s-4)^{-1}$ on
approaching the threshold from the finite-MSD regime.

Power-law restart therefore exhibits a hierarchy of long-time
thresholds with different physical origins. The threshold at $s=2$ is
set by renewal: only for $s>2$ is the mean waiting time finite, allowing
a normalized stationary age distribution and hence a normalized
stationary site-occupation profile. Higher thresholds arise from
transport moments of the ballistic restart-free walk. In particular,
the stationary MSD remains divergent for $2<s\leq4$ because rare long
restart-free intervals generate sufficiently large ballistic excursions,
and becomes finite only for $s>4$. Thus, for $1<s\leq2$ no normalized
stationary occupation profile exists, for $2<s\leq4$ the profile is
stationary but has an infinite MSD, and for $s>4$ both the profile and
its second moment are finite. More generally, for the ballistic
benchmark preparations considered here, the stationary $p$th absolute
spatial moment is finite only for $s>p+2$, see Appendix~\ref{app:powerlaw_msd} for details.

\section{Monitored first detection under sharp restart}
\label{sec:sharp_restart}

We now turn to a monitored first-detection problem in which a target
site is measured periodically and the walk is sharply restarted after
a fixed number of unsuccessful detection attempts
~\cite{yin_restart_2023,shukla_accelerated_2025}. This protocol differs
from the stochastic restart considered in the preceding sections.
There, the unitary dynamics is interrupted by restarts occurring at
random times, without intermediate measurements. Here, each null
measurement induces a projective backaction, whereas the restart itself
occurs deterministically.

Sharp restart has recently been shown to accelerate first detection in
the standard two-state DTQW, with the detection efficiency depending
sensitively on the monitoring interval, the restart time, and the
initial coin state~\cite{shukla_accelerated_2025}. Monitored recurrence
has also been studied for the corresponding one-parameter family of
three-state quantum walks~\cite{stefanak_monitored_2023}. Here, we
consider instead first detection at a spatially separated target under
deterministic sharp restart and investigate how the outcome depends on
the presence or absence of initial flat-band overlap.

\subsection{Monitoring and restart protocol}
\label{subsec:sharp_restart_protocol}

The walker is initialized at site $n_0$ with coin state
$\ket{\chi_0}$, and a detector is placed at
$n_D=n_0+\delta$, where $\delta$ denotes the displacement of the
target from the initial site. Since the measurement resolves position
but not the coin state, the detection and null-measurement projectors
are
\begin{align}
    \Pi_\delta
    &\equiv
    \mathbb I_3\otimes
    \ket{n_0+\delta}\bra{n_0+\delta},
    &
    Q_\delta
    &\equiv
    \mathbb I-\Pi_\delta .
    \label{eq:sharp_restart_projectors}
\end{align}
Measurements are performed every $\tau$ walk steps. The probability
that the walker is detected for the first time at the $j$th
measurement is
\begin{align}
    F_j(\ell,\delta,\tau|\chi_0)
    &\equiv
    \left\|
    \Pi_\delta U_\ell^\tau
    \left(
        Q_\delta U_\ell^\tau
    \right)^{j-1}
    \ket{\Psi(0)}
    \right\|^2 .
    \label{eq:sharp_restart_first_detection}
\end{align}
The factors of $Q_\delta$ incorporate the backaction of all preceding
null measurements.

Defining the survival probability after $j$ null measurements as
$ S_j \equiv \| (     Q_\delta U_\ell^\tau )^j \ket{\Psi(0)} \|^2$,
with $S_0=1$, the first-detection probability at the $j$th
measurement can be written as
$F_j= S_{j-1}-S_j$.
Indeed, using the orthogonal decomposition
$\Pi_\delta+Q_\delta=\mathbb I$ together with the unitarity of
$U_\ell^\tau$, one sees that 
$ S_{j-1} = \| U_\ell^\tau \left(Q_\delta U_\ell^\tau \right)^{j-1} \ket{\Psi(0)} \|^2 
=
\|  \Pi_\delta U_\ell^\tau \left(Q_\delta U_\ell^\tau \right)^{j-1} \ket{\Psi(0)} \! \|^2
+
\| Q_\delta U_\ell^\tau \left(Q_\delta U_\ell^\tau \right)^{j-1} \ket{\Psi(0)} \! \|^2
= F_j+S_j$.
The cumulative probability of detection by the
$N$th measurement, in the absence of restart, is therefore 
$P_{\rm det}(N) \equiv \sum_{j=1}^{N} F_j = S_0 - S_N $, which gives 
\begin{align}
    P_{\rm det}(N)
   = 1- \| \left(     Q_\delta U_\ell^\tau \right)^N \ket{\Psi(0)} \|^2 .
    \label{eq:sharp_restart_cumulative}
\end{align}

We impose a sharp restart after $r$ consecutive unsuccessful
measurements, where $r$ is a positive integer. Following the $r$th
null result, at time $t_r=r\tau$, the full walker-coin state is
restored to $\ket{\Psi(0)}$, and a new detection cycle begins.
Denoting by $P_r\equiv P_{\rm det}(r)$ the probability of detection
within one cycle, and writing $N=c r + m$ with $0\leq m <r$, the
cumulative detection probability under sharp restart is
\begin{align}
    P_{\rm det}^{(r)}(N)
    &=
    1-
    (1-P_r)^c
    \left[
        1-P_{\rm det}(m)
    \right].
    \label{eq:sharp_restart_cumulative_restart}
\end{align}
Hence, provided $P_r>0$, repeated restart drives the eventual
detection probability to unity.

Taking one walk step as the unit of time, the corresponding mean
first-detected-passage time (FDPT) is~\cite{bonomo_first_2021,yin_instability_2024}
\begin{align}
    \langle T\rangle_r
    &\equiv
    \tau
    \frac{r(1-P_r)}{P_r}
    +
    \tau
    \sum_{j=1}^{r}
    \frac{j F_j}{P_r}.
    \label{eq:sharp_restart_mean_fdpt}
\end{align}
The first term gives the mean time spent in complete failed cycles
before the successful one, whereas the second gives the mean detection
time within the final successful cycle.

In the numerical analysis below, the monitoring interval is fixed to
$\tau=1$, so that measurements are performed after every walk step.
For fixed $\ell$, $\delta$, and initial coin state, we determine the
optimal sharp-restart threshold $r^*$ by minimizing the mean FDPT over 
positive integers $r$, i.e.,
$r^*
\equiv
    \operatorname*{arg\,min}_{r\geq1}
    \langle T\rangle_r$.
The corresponding optimal restart time reads 
$t_r^*=r^*\tau$, and the minimum mean FDPT is defined as
$T_{\min}\equiv\langle T\rangle_{r^*}$.

\subsection{Flat-band control of first detection}
\label{subsec:sharp_restart_results}

We compare the flat-band-active state $\ket{s_\ell}$ with the 
flat-band-dark state $\ket{\chi_\perp}$ in the monitored first-detection problem.
Here, \emph{flat-band-dark}
refers specifically to the absence of overlap with the intrinsic flat
band of the LDTQW and should not be confused with a detector-dark
subspace generated by repeated 
measurements~\cite{friedman_quantum_2017,yin_restart_2023,wang_first_2024}.

\begin{figure}[tbp]
    \centering
    \includegraphics[width=0.82\linewidth]{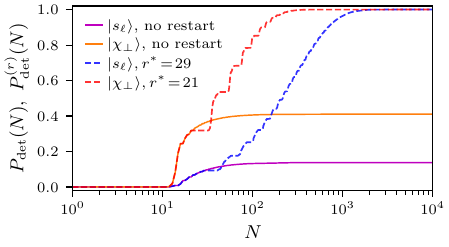}
    \caption{
    Cumulative detection probability for $\ell=2$, $\delta=10$, and
    $\tau=1$. Solid lines show the monitored dynamics without restart,
    Eq.~\eqref{eq:sharp_restart_cumulative}, for the flat-band-active state
    $\ket{s_\ell}$ and the flat-band-dark state $\ket{\chi_\perp}$. 
    Dashed lines show the corresponding dynamics 
    (Eq.~\eqref{eq:sharp_restart_cumulative_restart}) for  $\ket{s_\ell}$ and 
    $\ket{\chi_\perp}$ under
    optimal sharp restart, with $r^*=29$ and $r^*=21$, respectively.
    }
    \label{fig:sharp_restart_detection}
\end{figure}

Figure~\ref{fig:sharp_restart_detection} compares the cumulative
detection probabilities with and without sharp restart. In the
absence of restart, the flat-band-dark preparation reaches the distant
target substantially more efficiently than the flat-band-active
preparation. This contrast reflects their different initial spectral
content: $\ket{s_\ell}$ has finite overlap with the localized flat-band
sector, whereas $\ket{\chi_\perp}$ is initially supported entirely in
the dispersive sector. Repeated null measurements subsequently modify
the state through $Q_\delta$, so this initial spectral decomposition is
not generally preserved by the monitored dynamics.
Optimal sharp restart substantially accelerates the accumulation of
detection probability for both preparations, with the restarted
detection probabilities ultimately approaching unity.
The flat-band-dark preparation nevertheless reaches
high detection probabilities much earlier than the flat-band-active
preparation. Sharp restart therefore improves first-detection efficiency
without erasing its strong dependence on the initial spectral content.

\begin{figure}[tbp]
    \centering
    \includegraphics[width=\linewidth]{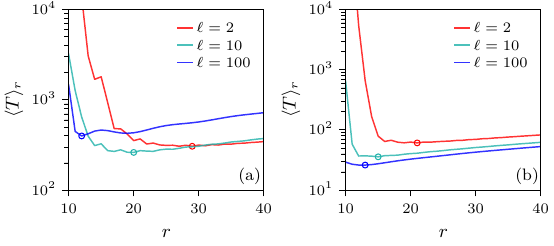}
    \caption{
    Mean FDPT $\langle T\rangle_r$ as a function of the 
    sharp-restart threshold $r$ at fixed $\delta=10$ and $\tau=1$.
    (a) Flat-band-active initial state $\ket{s_\ell}$.
    (b) Flat-band-dark initial state $\ket{\chi_\perp}$.
    Open circles mark the minima of the curves; their horizontal 
    coordinates give the corresponding optimal restart thresholds $r^*$.
    }
    \label{fig:fdpt_vs_r}
\end{figure}

The restart threshold $r$ itself provides an additional control parameter.
Figure~\ref{fig:fdpt_vs_r} shows the mean FDPT as a function of $r$ at
fixed $\delta=10$ and $\tau=1$ for several self-loop weights. For every
case shown, $\langle T\rangle_r$ possesses a finite minimum. Restarting
too early is inefficient for detection  because the probability $P_r$ of 
reaching the
target within one cycle is then small. In particular, for $\tau=1$ a
target at displacement $\delta$ cannot be detected before the
$\delta$th measurement. At the opposite extreme, an excessively long
cycle delays the initiation of a fresh search after unsuccessful
attempts. The optimal restart threshold therefore balances the time
required for propagation against the cost of remaining too long in an
unsuccessful cycle.

The optimal threshold $r^*$ and the corresponding minimum value of the 
mean FDPT depend strongly on the self-loop
weight and on the initial spectral preparation. For the flat-band-active
state, Fig.~\ref{fig:fdpt_vs_r}(a) already indicates a nonmonotonic
dependence of the optimal mean FDPT on $\ell$. In contrast,
Fig.~\ref{fig:fdpt_vs_r}(b) shows improved detection efficiency for the
flat-band-dark state as $\ell$ is increased, together with a shift of
the optimum toward shorter restart thresholds. The self-loop weight
therefore modifies both the detection efficiency and the sharp-restart
threshold that optimizes it.

\begin{figure}[tbp]
    \centering
    \includegraphics[width=\linewidth]{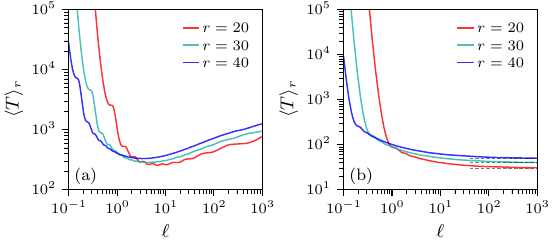}
    \caption{
    Mean FDPT as a function of the self-loop
    weight $\ell$ at fixed $\delta=10$ and $\tau=1$.
    (a) Flat-band-active initial state $\ket{s_\ell}$.
    (b) Flat-band-dark initial state $\ket{\chi_\perp}$.
    The gray dashed lines depict the limit~\eqref{eq:sharp_restart_dark_large_ell}.
    }
    \label{fig:fdpt_vs_ell}
\end{figure}

To isolate the role of the self-loop weight from the optimization over
$r$, we next hold the restart threshold fixed and vary $\ell$.
Figure~\ref{fig:fdpt_vs_ell}(a) shows that the active-state mean FDPT
has a broad minimum at an intermediate self-loop weight for each of the
restart thresholds considered. This behavior follows from two competing
limits of the LDTQW. The maximal dispersive group velocity,
$v_{\max}(\ell)=\sqrt{\ell/(\ell+2)}$, vanishes as $\ell\to0$, making
transport to a distant detector inefficient at sufficiently small
$\ell$. At the opposite extreme, as $\ell \to \infty$, the two benchmark coin states obey
$\ket{s_\ell} \to \ket{0}$ and
$\ket{\chi_\perp} \to (\ket{-1}+\ket{+1} ) / {\sqrt{2}}$.
The flat-band-active preparation therefore approaches the nonpropagating coin
state as $\ell$ increases, while the flat-band-dark preparation approaches an
equal superposition of the two moving components. The broad minimum in
Fig.~\ref{fig:fdpt_vs_ell}(a) consequently reflects a competition
between increasingly rapid dispersive transport and the suppression of
transport associated with the large-$\ell$ active preparation.

The contrasting dark-state behavior in
Fig.~\ref{fig:fdpt_vs_ell}(b) can be understood more explicitly from
the earliest possible detection event. For $\tau=1$ and a detector at
positive displacement $\delta$, detection at the earliest measurement,
$j=\delta$, requires a right-moving step at every walk step. For the two
benchmark states this gives (see Appendix~\ref{app:earliest_detection})
\begin{align}
    F_\delta^{\mathrm{act}}
    &=
    \frac{1}{\ell+2}
    \left(
        \frac{\ell}{\ell+2}
    \right)^{2(\delta-1)}, \label{eq:sharp_restart_earliest_detection_active} \\ 
    F_\delta^{\mathrm{dark}}
    &=
    \frac{\ell}{2(\ell+2)}
    \left(
        \frac{\ell}{\ell+2}
    \right)^{2(\delta-1)} .
    \label{eq:sharp_restart_earliest_detection_dark}
\end{align}
Thus $F_\delta^{\mathrm{act}}\sim\ell^{-1}$ in the flat-band-active sector, whereas
$F_\delta^{\mathrm{dark}}\to1/2$ in the flat-band-dark sector as $\ell\to\infty$.
The latter limit corresponds to a right-moving component that reaches
the detector at the earliest allowed time, while the remaining
left-moving component propagates away from the target. Consequently,
for fixed $r\geq\delta$,
\begin{align}
    \langle T\rangle_r^{\mathrm{dark}}
    &\to
    r+\delta,
    \qquad
    \ell\to\infty .
    \label{eq:sharp_restart_dark_large_ell}
\end{align}
This limiting behavior is consistent with the decreasing dark-state
curves in Fig.~\ref{fig:fdpt_vs_ell}(b). By contrast, the active
preparation approaches the nonpropagating state $\ket{0}$ as
$\ell\to\infty$, while its first-detection probability 
at the earliest accessible time vanishes as
$F_\delta^{\mathrm{act}}\sim\ell^{-1}$. Together, these limits account
for the eventual increase of $\langle T\rangle_r$ in
Fig.~\ref{fig:fdpt_vs_ell}(a).

The active-state curves also show weak oscillations superimposed on the
broad minimum. In Fig.~\ref{fig:fdpt_vs_ell}(a), $r$ is held fixed, so
these oscillations cannot result from changes in the optimal restart
threshold. At fixed $r$, the mean FDPT is determined by the finite
sequence $F_1,\ldots,F_r$. Varying $\ell$ changes the coin-scattering
amplitudes, i.e., the matrix elements of $C_\ell$ in
Eq.~\eqref{eq:coin_matrix}, as well as the dispersive quasienergy
phases entering the monitored propagator $Q_\delta U_\ell^\tau$, and
therefore modifies the coherent amplitudes contributing to the
first-detection probabilities $F_j$. We accordingly attribute the
oscillatory structure to finite-time coherent first-detection dynamics
rather than to distinct restart optima. This interpretation is
consistent with earlier work showing that quantum oscillations of the
first-detection probability can strongly affect sharp-restart
optimization~\cite{yin_instability_2024}. The flat-band-dark curves are
much smoother over the same range, because their purely dispersive
initial preparation involves a ballistic channel rather than
competition with a nonpropagating component.

Taken together, Figs.~\ref{fig:sharp_restart_detection}--%
\ref{fig:fdpt_vs_ell} show that the sharp-restart threshold $r$ and the
self-loop weight $\ell$ provide coupled controls of monitored first detection.
Their interplay depends strongly on the initial spectral content: the
flat-band-active preparation exhibits an optimal intermediate
self-loop regime, whereas the flat-band-dark preparation remains
efficient in the large-$\ell$ limit. Thus the presence or absence of
initial flat-band weight controls both the efficiency of
restart-assisted first detection and its response to the self-loop
weight.

\section{Conclusion}
\label{sec:conclusion}

We have studied stochastic and sharp restart in a one-dimensional
lackadaisical discrete-time quantum walk with coexisting flat and
dispersive bands. To separate the roles of these spectral sectors, we
used two benchmark initial coin states: a flat-band-active state with
finite flat-band overlap and a flat-band-dark state with zero flat-band
overlap.

Already without restart, the ballistic coefficient of the
flat-band-active state is nonmonotonic in $\ell$, reflecting the
competition between decreasing dispersive spectral weight and
increasing propagation speed within the dispersive sector.

For geometric stochastic restart, the local and global observables show
different weak-restart behavior. As the per-step restart probability
$q$ tends to zero, the stationary restart-site occupation of the
flat-band-active state approaches its restart-free intrinsic localized
value, whereas for the flat-band-dark state it vanishes as
$q\ln(1/q)$. The latter scaling follows from the $t^{-1}$ return
probability of the restart-free dispersive walk. The mean-squared
displacement exhibits a crossover from restart-free ballistic growth
to restart-induced confinement, with its stationary value scaling as
$q^{-2}$ in the same limit.

For power-law restart, the long-time behavior is controlled by the tail
exponent $s$. A normalized stationary site-occupation distribution
exists for $s>2$, while for the benchmark states considered here the
stationary $p$th absolute moment is finite only for $s>p+2$. In
particular, the stationary MSD is finite only for $s>4$. For
$1<s\leq2$, the mean waiting time to the next restart diverges and no
normalized stationary spatial distribution exists. Nevertheless, in
this same regime, at every fixed lattice site the flat-band-active
occupation approaches the intrinsic flat-band profile, while the
flat-band-dark occupation vanishes. Thus the localized and dispersive
sectors can have different local long-time limits even when the full
spatial distribution is not stationary.

We also considered monitored first detection under sharp restart, in
which the walk is reinitialized after a fixed number $r$ of consecutive
unsuccessful measurements. The number $r$ and the self-loop weight
jointly determine the mean first-detected-passage time. For fixed $r$,
the mean first-detected-passage time of the flat-band-active state has a
minimum at an intermediate self-loop weight. For the flat-band-dark
state, increasing the self-loop weight leads instead toward a simple
ballistic detection limit, with
$F_\delta^{\rm dark}\to1/2$ and
$\langle T\rangle_r^{\rm dark}\to r+\delta$ as $\ell\to\infty$ for
fixed $r\geq\delta$.

It is worth emphasizing which of these results are genuinely quantum. 
The flat band itself, and the finite-time oscillations underlying the 
first-detection probabilities, are coherent interference effects with 
no classical counterpart. The confinement scalings obtained under 
geometric and power-law restart, by contrast, follow once these 
quantum-mechanical return and transport probabilities are inserted 
into an otherwise classical renewal formalism; what is quantum there 
is the input, not the restart mechanism itself.

These results show that the response to restart depends strongly on the
spectral composition of the initial state. In the LDTQW, restart
therefore provides a way to distinguish localization inherited from the
flat band from confinement and modified first-detection behavior
generated by interruption of the dynamics. The same approach can be
extended to higher-dimensional quantum walks and lattice models with
coexisting localized and dispersive spectral sectors. It would also be
useful to extend the present analysis to partial
or subspace restart protocols, including spatially selective
reinitialization, to determine how incomplete restart modifies spectral
memory and the long-time dynamics.

Monitored quantum walks, in which unitary evolution
is interrupted stroboscopically by projective measurements and
first-detection statistics are recorded directly, have already been
implemented experimentally. Photonic recurrence measurements have
been realized on a time-multiplexed platform~\cite{schreiber_photons_2010,
nitsche_probing_2018}, while first-hitting-time statistics have been
measured on superconducting-qubit hardware using an IBM quantum
computer~\cite{wang_first_2024}. The sharp-restart protocol studied
here is therefore within reach of existing experimental platforms.

\section*{Acknowledgments}
The author acknowledges support from the MUR PRIN2022 project 
``Breakdown of ergodicity in classical and quantum many-body systems'' (BECQuMB) Grant No. 20222BHC9Z.

\section*{Data availability}
The data supporting the plots and results within this paper are 
available from the author upon reasonable request.

\appendix

\section{Evaluation of the flat-band Fourier integral}
\label{app:Gm_integral}

Using Eq.~\eqref{eq:lqw_flat_eigenvector} and $\mathcal P_0(k)=\ket{v_0(k)}\bra{v_0(k)}$, 
the flat-band kernel $K_{n-n_0}^{\mathrm{fb}}$ in Eq.~\eqref{eq:lqw_localized_kernel0} 
can be written explicitly in the matrix form:
\begin{widetext}
\begin{align}
    K_{n-n_0}^{\mathrm{fb}}
    =
    \begin{pmatrix}
        2G_{n-n_0} &
        \sqrt{\ell}(G_{n-n_0}+G_{n-n_0+1}) &
        2G_{n-n_0+1}
        \\
        \sqrt{\ell}(G_{n-n_0}+G_{n-n_0-1}) &
        \ell\left[
            G_{n-n_0}+\dfrac{G_{n-n_0+1}+G_{n-n_0-1}}{2}
        \right] &
        \sqrt{\ell}(G_{n-n_0}+G_{n-n_0+1})
        \\
        2G_{n-n_0-1} &
        \sqrt{\ell}(G_{n-n_0}+G_{n-n_0-1}) &
        2G_{n-n_0}
    \end{pmatrix},
    \label{eq:lqw_localized_kernel}
\end{align}
\end{widetext}
where we have
\begin{align}
    G_m(\ell)
    =
    \frac{1}{2\pi}
    \int_{-\pi}^{\pi}
    \frac{e^{ikm}}
    {4+\ell+\ell\cos k}\,dk,
    \quad
    m\in\mathbb{Z}.
    \label{eq:app_Gm_definition}
\end{align}
In writing Eq.~\eqref{eq:lqw_localized_kernel}, we have suppressed 
the argument $\ell$ from the function $G_m(\ell)$.
For $\ell=0$, one simply has $G_m(0) = \delta_{m,0} / 4$. Here, 
we evaluate this integral explicitly for $\ell>0$.
Since the denominator in Eq.~\eqref{eq:app_Gm_definition} is an even
function of $k$, while the integration interval is symmetric, one has
$G_{-m}(\ell) = G_m(\ell)$.
It is therefore sufficient to evaluate the integral for $m\geq0$ and
extend the result to arbitrary integer $m$ through the replacement
$m\rightarrow |m|$.

Introducing the complex variable $z=e^{ik}$ and considering the 
unit-circle contour traversed 
counterclockwise, Eq.~\eqref{eq:app_Gm_definition} may be written as
\begin{align}
    G_m(\ell)
    =
    \frac{1}{2\pi i}
    \oint_{|z|=1}
    \frac{2z^m}
    {\ell [  z^2+2(1+4/\ell)z+1 ] }
    \,dz .
    \label{eq:app_Gm_contour2}
\end{align}
The poles of the integrand are determined by the roots 
$z_{\pm}$ of $z^2 + 2 (1 + 4/\ell) z + 1 = 0$, which yields
\begin{align}
    z_{\pm}
    &=
    \frac{
        -(\ell+4)
        \pm
        2\sqrt{2(\ell+2)}
    }{\ell}, 
    \label{eq:app_poles}
\end{align}
Both roots are real and negative; they satisfy  $z_+z_-=1$. 
Explicitly, we have $-1<z_+<0$, and  $z_-<-1$, 
so that only $z_+$ lies inside the unit circle. 

Using the residue theorem, the integral~\eqref{eq:app_Gm_contour2} is computed as 
\begin{align}
    G_m(\ell) = \frac{2z_+^m} {\ell (z_+ - z_-)}  =
    \frac{ z_+^m }{ 2\sqrt{2(\ell+2) } },
    \quad
    m\geq0. 
    \label{eq:app_Gm_contour3}
\end{align}
Finally, using  
$G_{-m}=G_m$ and $z_+=-\eta(\ell)$, we obtain Eqs.~\eqref{eq:lqw_Gm} 
and \eqref{eq:lqw_eta} of the main text.

\section{Asymptotic return probability for the flat-band-dark state}
\label{app:dark_return}

The flat-band-dark state $\ket{\chi_\perp}$ given in Eq.~\eqref{eq:lqw_dark_state} 
satisfies  $\mathcal P_0(k)\ket{\chi_\perp}=0$ for all $k$. Hence, 
the restart-free coin state at the initial site $n=n_0$ contains only the 
dispersive contributions. From Eqs.~\eqref{eq:bare_amplitude_direct} 
and~\eqref{eq:lqw_dispersive_kernel}, we obtain
\begin{align}
    \ket{\psi_{n_0}(t)}
    =
    \int_{-\pi}^{\pi} \frac{dk}{2\pi}\,
    \left[
        e^{-i\omega(k)t}\mathcal P_+(k)
        +
        e^{i\omega(k)t}\mathcal P_-(k)
    \right]
    \ket{\chi_\perp} , 
    \label{eq:app_dark_return_amplitude}
\end{align}
where the frequencies $\pm \omega(k)$ are defined through the 
dispersion relation~\eqref{eq:dispersion}.

The long-time behavior of $\ket{\psi_{n_0}(t)}$ is governed 
by the stationary points of the phase appearing in the integrand of
Eq.~\eqref{eq:app_dark_return_amplitude}. Differentiating 
Eq.~\eqref{eq:dispersion} gives, whenever $\sin\omega(k)\neq0$,
\begin{align}
\omega'(k)
=
-\frac{\ell\sin k}
{(\ell+2)\sin\omega(k)}.
\label{eq:app_dark_omega_prime}
\end{align}
Thus regular stationary points can occur where $\sin k=0$, i.e., 
at $k=0$ and $k=\pi$ modulo $2\pi$.
At $k=0$, both the numerator and denominator vanish, so
Eq.~\eqref{eq:app_dark_omega_prime} is indeterminate. 
At $k=\pi$, instead,
$\sin\omega(\pi)=2\sqrt{2\ell}/(\ell+2)\neq0$, and therefore
$\omega'(\pi)=0$.
A second differentiation of Eq.~\eqref{eq:dispersion} gives the 
corresponding curvature 
\begin{align}
    \omega''(\pi)
    =
    \frac{\ell}
    {(\ell+2)\sin\omega(\pi)}
    =
    \frac{\sqrt{\ell}}{2\sqrt{2}}.
    \label{eq:app_dark_omega_second}
\end{align}

Let $y\equiv k-\pi$ and define
$\ket{a_\pm}\equiv\mathcal P_\pm(\pi)\ket{\chi_\perp}$.
Since $k=\pi$ and $k=-\pi$ represent the same point of the
Brillouin-zone circle, their endpoint neighborhoods combine into a
single neighborhood of the stationary point. For small $y$, one may write 
$
\omega(\pi+y) =
\omega(\pi)
+\frac{1}{2}\omega''(\pi)y^2
+O(y^3)$ 
and 
$\mathcal P_\pm(\pi+y)\ket{\chi_\perp} =
\ket{a_\pm}+O(y)$.
The quadratic phase varies by order unity when
$t y^2=O(1)$, so the leading contribution comes from
$|y|=O(t^{-1/2})$. For any fixed neighborhood $|y|<\delta$,
the corresponding scaled limits $\pm\delta\sqrt{t}$ therefore tend to
$\pm\infty$ as $t\to\infty$, allowing the local integral to be extended
over the real line at leading order. Equation~\eqref{eq:app_dark_return_amplitude}
then gives
\begin{align}
\ket{\psi_{n_0}(t)}
\sim
\frac{1}{2\pi}
\Bigg[
&e^{-i\omega(\pi)t}\ket{a_+}
\int_{-\infty}^{\infty}
dy\,
e^{-it\omega''(\pi)y^2/2}
\nonumber\\
+
&e^{i\omega(\pi)t}\ket{a_-}
\int_{-\infty}^{\infty}
dy\,
e^{it\omega''(\pi)y^2/2}
\Bigg].
\end{align}
Using the Fresnel integral  
$\int_{-\infty}^{\infty}
dy \, \exp(\pm iAy^2/2) =
\sqrt{{2\pi} / {A}}\,
\exp(\pm i\pi/4)$,
where $A$ is a positive real quantity,
we then obtain
\begin{align}
    \ket{\psi_{n_0}(t)}
    \sim
    \frac{1}
    {\sqrt{2\pi t\,\omega''(\pi)}}
    &\Big[
        e^{-i\omega(\pi)t-i\pi/4}\ket{a_+} \nonumber \\
        &+
        e^{i\omega(\pi)t+i\pi/4}\ket{a_-}
    \Big].
    \label{eq:app_dark_stationary_phase}
\end{align}
As the two dispersive eigenspaces are orthogonal, i.e., 
$\mathcal P_+(\pi)\mathcal P_-(\pi)=0$, one has
$\braket{a_+}{a_-}=0$. Moreover, since the flat-band projection of
$\ket{\chi_\perp}$ vanishes and the state is normalized, we have
$\braket{a_+}{a_+}+\braket{a_-}{a_-}=1$.
Taking the squared norm of
Eq.~\eqref{eq:app_dark_stationary_phase} therefore yields
\begin{align}
    P^{(0)}(n_0,t|n_0,\chi_\perp)
    &\sim
    \frac{1}
    {2\pi t\,\omega''(\pi)} ,
    \label{eq:app_dark_return_asymptotic}
\end{align}
which along with Eq.~\eqref{eq:app_dark_omega_second} gives 
Eq.~\eqref{eq:lqw_dark_return} of the main text.

\section{Ballistic limit of the restart-free walker}
\label{app:restart_free_msd_ballistic}

We derive here the long-time behavior of spatial moments of the restart-free
LDTQW. Consider a walker initially localized at site $n_0$ with a
normalized coin state $\ket{\chi_0}$, i.e., 
$\ket{\Psi(0)} =
    \ket{\chi_0}\otimes\ket{n_0}$, 
as given in Eq.~\eqref{eq:lqw_initial_state_psi0}.
In Fourier space the evolution is given by
\begin{align}
    \ket*{\widetilde{\psi}(k,t)}
    &=
    e^{-ikn_0}
    [U_\ell(k)]^t\ket{\chi_0}.
    \label{eq:app_fourier_evolution}
\end{align}

Let $n_t$ denote the lattice position of the walker at time $t$, so that the 
displacement is given by $X_t \equiv 
    n_t-n_0$.
Since the
initial state is localized at $n_0$, repeated application of the 
shift operator~\eqref{eq:lqw_shift}
restricts the support after $t$ steps to
$n_0-t \leq n_t \leq n_0+t$, and hence $|X_t| \leq t$.
We define the corresponding scaled displacement as
\begin{align}
    V_t \equiv X_t/t , 
    \qquad
    |V_t|
    \leq
    1.
    \label{eq:app_scaled_displacement_bound}
\end{align}
For $p>0$, the $p$th absolute displacement moment is
\begin{align}
    M_p^{(0)}(t)
    &\equiv
    \sum_{n=-\infty}^{\infty} 
    |n-n_0|^p
    P^{(0)}(n,t|n_0,\chi_0).
    \label{eq:app_absolute_moment}
\end{align}
To determine its leading long-time behavior, we first obtain the
limiting distribution of $V_t$.

The characteristic function of $V_t$ is given by
\begin{align}
    \varphi_t(\xi)
    &\equiv
    \left\langle e^{i\xi V_t}\right\rangle
    =
    \sum_{n=-\infty}^{\infty}
    e^{i\xi(n-n_0)/t}
    P^{(0)}(n,t|n_0,\chi_0)
    \nonumber\\
    &=
    e^{-i\xi n_0/t}
    \sum_{n=-\infty}^{\infty}
    e^{i\xi n/t}
    \braket{\psi_n(t)}{\psi_n(t)} \nonumber \\
    &=
    \frac{e^{-i\xi n_0/t}}{2\pi}
    \int_{-\pi}^{\pi}dk\,
    \braket*{\widetilde{\psi}(k+\xi/t,t)}
           {\widetilde{\psi}(k,t)}.
    \label{eq:app_scaled_characteristic}
\end{align}
Substituting Eq.~\eqref{eq:app_fourier_evolution} into
Eq.~\eqref{eq:app_scaled_characteristic} cancels the phase associated with the
initial position and yields
\begin{align}
    \varphi_t(\xi)
    =
    \frac{1}{2\pi}
    \int_{-\pi}^{\pi}dk\,
    \bra{\chi_0}
    [U_\ell(k+\xi/t)]^{\dagger t}
    [U_\ell(k)]^t
    \ket{\chi_0} , 
    \label{eq:app_scaled_characteristic_reduced}
\end{align}
which makes explicit that the displacement statistics are independent
of the initial position $n_0$.

Using spectral projectors $\mathcal P_\sigma(k)$ with $\sigma\in\{0,+,-\}$, 
introduced in Eq.~\eqref{eq:lqw_spectral_decomposition}, one may 
define the spectral weights 
$w_\sigma(k) \equiv 
    \bra{\chi_0}
    \mathcal P_\sigma(k)
    \ket{\chi_0}$.
For every $k\neq0$, the spectral projectors are smooth in $k$ and
orthogonal. Hence, for fixed $\xi$ and $t\to \infty $, we have $\mathcal P_\sigma(k+\xi/t)
    \mathcal P_{\sigma'}(k)
    \to
    \delta_{\sigma\sigma'}
    \mathcal P_\sigma(k)$.
The dispersive phases, in the same limit, give 
    $t\left[
        \omega(k+\xi/t)-\omega(k)
    \right]
    \to 
    \xi v(k)$,
where  $v(k) = {d\omega(k)} / {dk}$.
The only exception is the isolated dispersive-band degeneracy at
$k=0$, which has zero measure in the quasimomentum integral. Moreover,
by unitarity, one has 
$|
\bra{\chi_0}
[U_\ell(k+\xi/t)]^{\dagger t}
[U_\ell(k)]^t
\ket{\chi_0}|\leq1$
uniformly in $k$ and $t$. Dominated convergence therefore allows the
limit $t\to\infty$ to be taken inside the quasimomentum integral.
Consequently, using Eqs.~\eqref{eq:lqw_spectral_decomposition} 
and~\eqref{eq:app_scaled_characteristic_reduced}, we obtain the 
long-time limit $\varphi_t(\xi) \to \varphi(\xi)$, where 
\begin{align}
    \varphi(\xi)
    =
    \frac{1}{2\pi}
    \int_{-\pi}^{\pi}dk\,
    \big[
        w_0(k)
        &+
        w_+(k)e^{i\xi v(k)} \nonumber \\
        &+
        w_-(k)e^{-i\xi v(k)}
    \big].
    \label{eq:app_velocity_characteristic}
\end{align}

Equation~\eqref{eq:app_velocity_characteristic} gives the pointwise
limit of the characteristic functions
of the scaled displacement $V_t$. To infer convergence in distribution,
we first note that
\begin{align}
\varphi(0)
&=
\frac{1}{2\pi}
\int_{-\pi}^{\pi}dk\,
\left[
w_0(k)+w_+(k)+w_-(k)
\right]
=1,
\end{align}
where the completeness of the spectral projectors has been used.
Moreover, $\varphi(\xi)$ is continuous at $\xi=0$, since the factors
$e^{\pm i\xi v(k)}$ are continuous in $\xi$, while
the absolute value of the integrand is bounded by
$w_0(k)+w_+(k)+w_-(k)=1$. Hence, by L\'evy's continuity theorem, 
$V_t$ converges in distribution to a random variable $V$ with characteristic function
$\varphi(\xi)$ such that
\begin{align}
V_t
\xrightarrow[t\to\infty]{\mathrm d}
V.
\label{eq:app_scaled_displacement_weak_limit}
\end{align}
Thus,  Eq.~\eqref{eq:app_velocity_characteristic} determines the
long-time probability distribution of the scaled displacement. 
The functional form of $\varphi(\xi)$ admits a direct 
interpretation in terms of the band group velocities.

The distribution of $V$ can be read directly from this characteristic
function. For every quasimomentum $k\neq 0$, the three spectral bands carry
the group velocities
$0$, $+v(k)$, and $-v(k)$,
with corresponding spectral weights
$w_0(k)$, $w_+(k)$, and $w_-(k)$. The quasimomentum itself is averaged
over the Brillouin zone with measure $dk/(2\pi)$. Equivalently, the
limiting probability measure may be viewed as a measure on the joint
space $(k,\sigma)$ defined by
\begin{align}
d\mu(k,\sigma) \equiv 
\frac{dk}{2\pi}\,w_\sigma(k),
\qquad
\sigma=0,+,-,
\end{align}
together with the map
\begin{align}
(k,0)&\mapsto 0,
&
(k,+)&\mapsto v(k),
&
(k,-)&\mapsto -v(k).
\end{align}
Indeed, since
$w_\sigma(k)\geq0$ and
$w_0(k)+w_+(k)+w_-(k)=1$, this measure is normalized, i.e., 
$
\sum_{\sigma=0,+,-}
\int_{-\pi}^{\pi}
d\mu(k,\sigma)
=
1$.
Therefore, for any measurable function $f(V)$ for which the 
expectation exists, one may write 
\begin{align}
\langle f(V)\rangle
=
\frac{1}{2\pi}
\int_{-\pi}^{\pi}dk\,
&\Big[
w_0(k)f(0)
+w_+(k)f(v(k)) \nonumber \\
&+w_-(k)f(-v(k))
\Big].
\label{eq:app_velocity_average}
\end{align}
Choosing $f(V)=e^{i\xi V}$ in
Eq.~\eqref{eq:app_velocity_average} reproduces 
Eq.~\eqref{eq:app_velocity_characteristic}. Thus the finite-time
scaled displacement $V_t=X_t/t$ does not equal a group velocity;
rather, its long-time distribution is the spectrally weighted
distribution of the band group velocities over the entire Brillouin
zone.

We now apply this result to the spatial moments. Since $X_t=tV_t$,
one has
$M_p^{(0)}(t)/t^p=\langle |V_t|^p\rangle$.
Equation~\eqref{eq:app_scaled_displacement_bound} gives
$|V_t|\leq1$ for every $t$, so all $V_t$ have support in the common
compact interval $[-1,1]$; their weak limit $V$ is supported there as well. 
Since $|x|^p$ is continuous and bounded in this interval, the
convergence~\eqref{eq:app_scaled_displacement_weak_limit} implies, for $p>0$,
\begin{align}
\lim_{t\to\infty}
\frac{M_p^{(0)}(t)}{t^p}
=
\lim_{t\to\infty}
\left\langle |V_t|^p\right\rangle
=
\left\langle |V|^p\right\rangle.
\end{align}
Finally, setting $f(V)=|V|^p$ in
Eq.~\eqref{eq:app_velocity_average}, we obtain 
\begin{align}
\lim_{t\to\infty}
\frac{M_p^{(0)}(t)}{t^p}
&=
\mathcal B_p(\ell,\chi_0),
\nonumber\\
\mathcal B_p(\ell,\chi_0)
&=\frac{1}{2\pi}
\int_{-\pi}^{\pi}dk\,
|v(k)|^p
\left[
w_+(k)+w_-(k)
\right] \nonumber \\
&=
\frac{1}{2\pi}
\int_{-\pi}^{\pi}dk\,
|v(k)|^p
\left[
1-
\bra{\chi_0}
\mathcal P_0(k)
\ket{\chi_0}
\right],
\label{eq:app_ballistic_moment_limit}
\end{align}
where, in obtaining the last equality, we have used
$\mathcal P_0(k)+\mathcal P_+(k)+\mathcal P_-(k)=\mathbb{I}_3$. 
The flat band does not contribute to $\mathcal B_p(\ell,\chi_0)$ because its group
velocity vanishes.
Equation~\eqref{eq:app_ballistic_moment_limit} implies that 
for $\mathcal B_p(\ell,\chi_0)>0$, one has 
\begin{align}
    M_p^{(0)}(t)
    \sim
    \mathcal B_p(\ell,\chi_0)t^p,
    \qquad
    t\to\infty .
    \label{eq:app_ballistic_moment_asymptotic}
\end{align}

For the MSD, i.e., $p=2$, we write
$\mathcal B(\ell,\chi_0)\equiv\mathcal B_2(\ell,\chi_0)$ so 
that Eqs.~\eqref{eq:app_ballistic_moment_asymptotic} 
and~\eqref{eq:app_ballistic_moment_limit} reduce to 
Eqs.~\eqref{eq:comp_bare_ballistic_r0} and~\eqref{eq:comp_ballistic_coeff} 
of the main text, respectively.
From Eq.~\eqref{eq:lqw_group_velocity},
\begin{align}
    v^2(k)
    =
    1-
    \frac{4}
         {\ell+4+\ell\cos k}.
    \label{eq:app_velocity_squared}
\end{align}
We now evaluate Eq.~\eqref{eq:app_ballistic_moment_limit} for the two
benchmark coin states used in the main text.

For the flat-band-active state $\ket{s_\ell}$,
\begin{align}
    \bra{s_\ell}
    \mathcal P_0(k)
    \ket{s_\ell}
    =
    \frac{(\ell+2)(1+\cos k)}
         {\ell+4+\ell\cos k}.
    \label{eq:app_active_overlap}
\end{align}
Combining this expression with
Eqs.~\eqref{eq:app_ballistic_moment_limit} and~\eqref{eq:app_velocity_squared} gives
\begin{align}
    \mathcal B(\ell,s_\ell)
    =
    \frac{\ell}{\pi}
    \int_{-\pi}^{\pi}
    dk\,
    \frac{1-\cos^2 k}
         {[\ell+4+\ell\cos k]^2} ,
    \label{eq:app_Bactive_integral}
\end{align}
which evaluates to Eq.~\eqref{eq:comp_benchmark_B_active}. 
On the other hand, for the flat-band-dark state $\ket{\chi_\perp}$,
$\mathcal P_0(k)\ket{\chi_\perp}=0$ for all $k$. 
Therefore, from Eqs.~\eqref{eq:app_ballistic_moment_limit} 
and~\eqref{eq:app_velocity_squared}, we obtain 
\begin{align}
    \mathcal B(\ell,\chi_\perp)
    &=
    1-
    \frac{4}{2\pi}
    \int_{-\pi}^{\pi}
    \frac{dk}
         {\ell+4+\ell\cos k} ,
    \label{eq:app_Bdark_integral}
\end{align}
which yields Eq.~\eqref{eq:comp_benchmark_B_dark} of the main text.

\section{Weak-restart MSD for geometric restart}
\label{app:geometric_msd}

The exact finite-time MSD for geometric restart is given in Eq.~\eqref{eq:geom_msd_transient} of the main text.
Let us consider the weak-restart scaling limit
$q\to0^+$ and $t\to\infty$ with
$x=qt$ fixed.
In this limit, the relevant ages are large, so
that the restart-free MSD scales as $M_2^{(0)}(t)\sim\mathcal B(\ell,\chi_R)  t^2$, while
$(1-q)^a\sim e^{-qa}$. 
The no-restart contribution, i.e., the first term on the right-hand side of Eq.~\eqref{eq:geom_msd_transient} then becomes
\begin{align}
    (1-q)^tM_2^{(0)}(t)
    &\sim
    \frac{\mathcal B (\ell,\chi_R)}{q^2}
    x^2e^{-x}.
    \label{eq:app_geom_no_restart_scaling}
\end{align}
Introducing a scaled age $y=qa$ turns the renewal sum, i.e., the second term on the right-hand side of Eq.~\eqref{eq:geom_msd_transient} into an integral:
\begin{align}
    q\sum_{a=0}^{t-1}
    (1-q)^aM_2^{(0)}(a)
    &\sim
    \frac{\mathcal B(\ell,\chi_R)}{q^2}
    \int_0^x dy\,y^2e^{-y}.
    \label{eq:app_geom_renewal_scaling}
\end{align}
Combining Eqs.~\eqref{eq:app_geom_no_restart_scaling} and
\eqref{eq:app_geom_renewal_scaling}, we obtain from Eq.~\eqref{eq:geom_msd_transient} that 
\begin{align}
    M_{2,\mathrm g}(t)
    &\sim
    \frac{2\mathcal B (\ell, \chi_R)}{q^2}
    \left[
        1-(1+x)e^{-x}
    \right],
    \quad
    x=qt,
    \label{eq:app_geom_crossover_x}
\end{align}
which is 
Eq.~\eqref{eq:geom_msd_crossover} of the main text.

\section{MSD and stationary moment hierarchy for power-law restart}
\label{app:powerlaw_msd}

Applying the renewal transform
\eqref{eq:general_renewal_transform} to the power-law restart gives  the corresponding MSD generating function 
\begin{align}
\widetilde M_{2,\mathrm{pl}}(z)
&=
\frac{N_s(z)}{D_s(z)},
\end{align}
where we have 
\begin{align}
N_s(z)
&\equiv 
\sum_{a=1}^\infty
\zeta(s,a+1)M_2^{(0)}(a)z^a,
\\
D_s(z)
&\equiv
\zeta(s)-\operatorname{Li}_s(z).
\label{eq:app_powerlaw_msd_generating}
\end{align}
For large $a$,
$\zeta(s,a+1)\sim a^{1-s}/(s-1)$, while
$M_2^{(0)}(a)\sim\mathcal B(\ell, \chi_R) a^2$. Thus, the $a$-dependent terms entering the sum 
$N_s(z)$ scale as $a^{3-s}$. Consequently, the corresponding 
asymptotics as $z\to1^-$ are obtained as 
\begin{align}
N_s(z)
&\sim
\frac{\mathcal B(\ell, \chi_R) \, \Gamma(4-s)}{s-1}
(1-z)^{s-4},
&&1<s<4, \label{eq:app_powerlaw_numerator} \\
N_4(z)
&\sim
\frac{\mathcal B(\ell, \chi_R)}{3}
\ln\frac{1}{1-z},
&&s=4.
\end{align}
For $s>4$, $N_s(1)$ is finite.

The standard expansion of the polylogarithm near $z=1$ gives
\begin{align}
D_s(z)
&\sim
-\Gamma(1-s)(1-z)^{s-1},
&&1<s<2, \\
D_s(z)
&\sim
(1-z)\ln\frac{1}{1-z}, 
&&s=2,\\
D_s(z)
&\sim
\zeta(s-1)(1-z),
&&s>2.
\label{eq:app_powerlaw_denominator}
\end{align}

The same asymptotics also justify the age-divergence statement used in
the main text. Since
$\widetilde u_{\mathrm{pl}}(z)=\zeta(s)/D_s(z)$,
Eq.~\eqref{eq:app_powerlaw_denominator} gives
$u_j^{(\mathrm{pl})}\sim C_s j^{s-2}$ as $j\to\infty$ for
$1<s<2$, with $C_s>0$, and
$u_j^{(\mathrm{pl})}\sim \zeta(2)/\ln j$ as $j\to\infty$ at
$s=2$. Hence $u_j^{(\mathrm{pl})}\to0$ throughout
$1<s\leq2$. Recall that $A_t$ is the number of steps elapsed since
the most recent restart at observation time $t$. For any fixed finite
integer $M$, the event $A_t\leq M$ means that the most recent restart
occurred within the last $M$ steps, and the exact age distribution
therefore gives
\begin{align}
\Pr(A_t\leq M)
&=
\sum_{a=0}^{M}
u_{t-a}^{(\mathrm{pl})}\Phi_a^{(\mathrm{pl})}
\longrightarrow 0,
\qquad t\to\infty ,
\label{eq:app_powerlaw_age_divergence}
\end{align}
with $M$ held fixed. Indeed, for every fixed $a\leq M$,
$t-a\to\infty$ and hence $u_{t-a}^{(\mathrm{pl})}\to0$, while the
sum contains only finitely many terms. Thus $A_t\to\infty$ in
probability for $1<s\leq2$. Since, at every fixed lattice site $n$,
the restart-free occupation probability $P^{(0)}(n,a|n_R,\chi_R)$ 
approaches its flat-band contribution $P_{\mathrm{fb}}^{(0)}(n|n_R,\chi_R)$ 
as the age $a\to\infty$, this establishes the pointwise limit in
Eq.~\eqref{eq:comp_powerlaw_pointwise}; it does not imply convergence
to a normalized stationary site-occupation distribution.

Combining Eqs.~\eqref{eq:app_powerlaw_numerator}--~\eqref{eq:app_powerlaw_denominator} 
gives the following leading order behaviors of the MSD generating 
function, in the limit $z\to1$:
\begin{align}
\widetilde M_{2,\mathrm{pl}}(z)
\!\sim \!
\begin{cases}
\mathcal B(\ell, \chi_R)(2-s)(3-s)(1-z)^{-3},
&1\!<\!s\!<\!2,\\[1mm]
\displaystyle
\mathcal B(\ell, \chi_R)
\frac{(1-z)^{-3}}
{\ln[1/(1-z)]},
&s=2,\\[3mm]
\displaystyle
\frac{\mathcal B(\ell, \chi_R)\Gamma(4-s)}
{(s-1)\zeta(s-1)}
(1-z)^{-(5-s)},
&2\!<\!s\!<\!4,\\[3mm]
\displaystyle
\frac{\mathcal B(\ell, \chi_R)}{3\zeta(3)}
\frac{\ln[1/(1-z)]}{1-z},
&s=4.
\end{cases}
\label{eq:app_powerlaw_singular_summary}
\end{align}
Applying the standard Tauberian coefficient-transfer results for these
algebraic and logarithmic singularities yields the four long-time MSD
regimes reported in Eq.~\eqref{eq:comp_powerlaw_scaling}. 

For $s>4$, the sum $N_s(1)= 
\sum_{a=1}^\infty
\zeta(s,a+1)M_2^{(0)}(a)$ is finite, so the stationary-age
representation yields
$M_{2,\mathrm{pl}}^{\mathrm{st}}=N_s(1)/\zeta(s-1)$.
Replacing the restart-free MSD by its leading ballistic form
$M_2^{(0)}(a)\sim\mathcal B(\ell,\chi_R)a^2$ throughout the stationary
renewal sum gives
\begin{align}
   M_{2,\mathrm{pl}}^{\mathrm{st}} 
   &\approx \frac{\mathcal B(\ell, \chi_R)}{\zeta(s-1)} \sum_{a=1}^{\infty}
\zeta(s,a+1) a^2 \nonumber \\
   &=\frac{\mathcal B(\ell, \chi_R)}{\zeta(s-1)}\sum_{a=0}^{\infty} \sum_{k=1}^{\infty} \frac{a^2}{(a+k)^s} \nonumber \\
   &=\frac{\mathcal B(\ell, \chi_R)}{\zeta(s-1)} \sum_{n=1}^{\infty} \frac{1}{n^s} \sum_{a=0}^{n-1} a^2.
\end{align}
Here, the $a=0$ term vanishes identically, and the last line follows by
setting $n=a+k$ and interchanging the absolutely convergent sums.
Evaluating the finite sum over $a$ then gives
Eq.~\eqref{eq:comp_powerlaw_scaling_msd_steady} of the main text.

The same argument extends directly to arbitrary absolute moments.
From Eq.~\eqref{eq:app_ballistic_moment_asymptotic},
$M_p^{(0)}(a)\sim\mathcal B_p(\ell,\chi_R)a^p$, while the stationary
power-law age distribution behaves as
$\pi_a^{(\mathrm{pl})} \sim a^{1-s}$. Hence, the large-age contribution to the
stationary $p$th absolute moment scales as $a^{p+1-s}$. The
corresponding age sum converges if and only if
$s > p+2$.
Thus the finite-MSD condition $s>4$ is the $p=2$ member of the more
general stationary-moment hierarchy.

\section{First detection at the earliest accessible time}
\label{app:earliest_detection}

For monitoring after every walk step, $\tau=1$, with the detector located at
$n_D=n_0+\delta$ and $\delta>0$, the earliest possible detection occurs at the
$\delta$th measurement, since the walker can advance by at most one lattice site
per step. The detector is therefore kinematically inaccessible at all earlier
measurement times, so the preceding null measurements do not affect this event.
Detection at $t=\delta$ is possible only if the walker undergoes a right-moving
shift at every step, and the corresponding first-detection probability is
$F_\delta$.

In the coin basis $\{\ket{-1},\ket{0},\ket{+1}\}$, the amplitude for
this unique path is therefore
\begin{align}
    \mathcal S_\delta(\chi_0)
    &=
    \bra{+1}C_\ell\ket{\chi_0}
    \left(
        \bra{+1}C_\ell\ket{+1}
    \right)^{\delta-1}.
    \label{eq:app_earliest_general}
\end{align}

Using $C_\ell=2\ket{s_\ell}\bra{s_\ell}- \mathbb{I}_3$ gives
$\bra{+1}C_\ell\ket{+1}
=
-{\ell}/ (\ell+2)$.
For the flat-band-active preparation, $C_\ell\ket{s_\ell}=\ket{s_\ell}$,
and hence
$\bra{+1}C_\ell\ket{s_\ell}
= 1 / ({\sqrt{\ell+2}})$.
The flat-band-dark state satisfies
$\braket{s_\ell}{\chi_\perp}=0$ and therefore
$C_\ell\ket{\chi_\perp}=-\ket{\chi_\perp}$ and 
$\bra{+1}C_\ell\ket{\chi_\perp}
= - \sqrt{\ell} / \sqrt{2(\ell+2)}$. 
The first-detection probability at the earliest accessible time for a
general initial coin state is therefore
$F_\delta(\ell,\delta,1|\chi_0)
=|\mathcal S_\delta(\chi_0)|^2$.
For $\ket{\chi_0}=\ket{s_\ell}$ and
$\ket{\chi_0}=\ket{\chi_\perp}$, this yields
Eqs.~\eqref{eq:sharp_restart_earliest_detection_active} and
\eqref{eq:sharp_restart_earliest_detection_dark}, respectively.

\bibliographystyle{unsrt}
\bibliography{library} 

@article{acharya_tight-binding_2023,
  title = {Tight-Binding Model Subject to Conditional Resets at Random Times},
  author = {Acharya, A. and Gupta, S.},
  year = 2023,
  journal = {Phys. Rev. E},
  volume = {108},
  number = {6},
  pages = {064125},
  doi = {10.1103/PhysRevE.108.064125},
  langid = {english}
}

@article{ambainis_quantum_2007,
  title = {Quantum Walk Algorithm for Element Distinctness},
  author = {Ambainis, A.},
  year = 2007,
  journal = {SIAM J. Comput.},
  volume = {37},
  number = {1},
  pages = {210--239},
  doi = {10.1137/S0097539705447311},
  langid = {english}
}

@article{bonomo_first_2021,
  title = {First Passage under Restart for Discrete Space and Time: {{Application}} to One-Dimensional Confined Lattice Random Walks},
  author = {Bonomo, O. L. and Pal, A.},
  year = 2021,
  journal = {Phys. Rev. E},
  volume = {103},
  number = {5},
  pages = {052129},
  publisher = {American Physical Society},
  doi = {10.1103/PhysRevE.103.052129},
}

@article{carollo_stochastic_2026,
  title = {Stochastic Resetting Induces Quantum Non-{{Markovianity}}},
  author = {Carollo, F. and Wald, S.},
  year = 2026,
  journal = {arXiv:2601.13367},
  eprint = {2601.13367},
  primaryclass = {quant-ph},
  publisher = {arXiv},
  doi = {10.48550/arXiv.2601.13367},
  archiveprefix = {arXiv},
}

@article{chandrashekar_optimizing_2008,
  title = {Optimizing the Discrete Time Quantum Walk Using a {{SU}}(2) Coin},
  author = {Chandrashekar, C. M. and Srikanth, R. and Laflamme, R.},
  year = 2008,
  journal = {Phys. Rev. A},
  volume = {77},
  number = {3},
  pages = {032326},
  doi = {10.1103/PhysRevA.77.032326},
  langid = {english},
}

@article{chelminiak_discrete-time_2025,
  title = {Discrete-Time Walk on One-Dimensional Lattice under Stochastic Resetting: {{Advantage}} of Quantum over Classical Scenario},
  shorttitle = {Discrete-Time Walk on One-Dimensional Lattice under Stochastic Resetting},
  author = {Che{\l}miniak, P. and W{\'o}jcik, J. and W{\'o}jcik, A.},
  year = 2025,
  journal = {Phys. Rev. E},
  volume = {111},
  number = {4},
  pages = {044143},
  doi = {10.1103/PhysRevE.111.044143},
  langid = {english},
}

@article{childs_universal_2009,
  title = {Universal Computation by Quantum Walk},
  author = {Childs, A. M.},
  year = 2009,
  journal = {Phys. Rev. Lett.},
  volume = {102},
  number = {18},
  pages = {180501},
  doi = {10.1103/PhysRevLett.102.180501},
  langid = {english},
}

@article{das_discrete_2022,
  title = {Discrete Space-Time Resetting Model: Application to First-Passage and Transmission Statistics},
  shorttitle = {Discrete Space-Time Resetting Model},
  author = {Das, D. and Giuggioli, L.},
  year = 2022,
  journal = {J. Phys. A: Math. Theor.},
  volume = {55},
  number = {42},
  pages = {424004},
  doi = {10.1088/1751-8121/ac9765},
}

@article{das_quantum_2022,
  title = {Quantum Random Walk and Tight-Binding Model Subject to Projective Measurements at Random Times},
  author = {Das, D. and Gupta, S.},
  year = 2022,
  journal = {J. Stat. Mech.: Theory Exp.},
  volume = {2022},
  number = {3},
  pages = {033212},
  doi = {10.1088/1742-5468/ac5dc0},
}

@article{das_quantum_2022-1,
  title = {Quantum Unitary Evolution Interspersed with Repeated Non-Unitary Interactions at Random Times: The Method of Stochastic {{Liouville}} Equation, and Two Examples of Interactions in the Context of a Tight-Binding Chain},
  shorttitle = {Quantum Unitary Evolution Interspersed with Repeated Non-Unitary Interactions at Random Times},
  author = {Das, D. and Dattagupta, S. and Gupta, S.},
  year = 2022,
  journal = {J. Stat. Mech.: Theory Exp.},
  volume = {2022},
  number = {5},
  pages = {053101},
  doi = {10.1088/1742-5468/ac6256},
}

@article{dattagupta_stochastic_2022,
  title = {Stochastic Resets in the Context of a Tight-Binding Chain Driven by an Oscillating Field},
  author = {Dattagupta, S. and Das, D. and Gupta, S.},
  year = 2022,
  journal = {J. Stat. Mech.: Theory Exp.},
  volume = {2022},
  number = {10},
  pages = {103210},
  doi = {10.1088/1742-5468/ac98c0},
}

@article{dhar_detection_2015,
  title = {Detection of a Quantum Particle on a Lattice under Repeated Projective Measurements},
  author = {Dhar, S. and Dasgupta, S. and Dhar, A. and Sen, D.},
  year = 2015,
  journal = {Phys. Rev. A},
  volume = {91},
  number = {6},
  pages = {062115},
  doi = {10.1103/PhysRevA.91.062115},
  langid = {english},
}

@article{dhar_quantum_2015,
  title = {Quantum Time of Arrival Distribution in a Simple Lattice Model},
  author = {Dhar, S. and Dasgupta, S. and Dhar, A.},
  year = 2015,
  journal = {J. Phys. A: Math. Theor.},
  volume = {48},
  number = {11},
  pages = {115304},
  doi = {10.1088/1751-8113/48/11/115304},
}

@article{dubey_quantum_2023,
  title = {Quantum Resetting in Continuous Measurement Induced Dynamics of a Qubit},
  author = {Dubey, V. and Chetrite, R. and Dhar, A.},
  year = 2023,
  journal = {J. Phys. A: Math. Theor.},
  volume = {56},
  number = {15},
  pages = {154001},
  doi = {10.1088/1751-8121/acc290},
}

@article{erdos_property_1949,
  title = {A Property of Power Series with Positive Coefficients},
  author = {Erd{\"o}s, P. and Feller, W. and Pollard, H.},
  year = 1949,
  journal = {Bull. Amer. Math. Soc.},
  volume = {55},
  number = {2},
  pages = {201--204},
  doi = {10.1090/S0002-9904-1949-09203-0},
  langid = {english}
}

@article{eule_non-equilibrium_2016,
  title = {Non-Equilibrium Steady States of Stochastic Processes with Intermittent Resetting},
  author = {Eule, S. and Metzger, J. J.},
  year = 2016,
  journal = {New J. Phys.},
  volume = {18},
  number = {3},
  pages = {033006},
  publisher = {IOP Publishing},
  doi = {10.1088/1367-2630/18/3/033006},
  langid = {english},
}

@article{evans_diffusion_2011,
  title = {Diffusion with Stochastic Resetting},
  author = {Evans, M. R. and Majumdar, S. N.},
  year = 2011,
  journal = {Phys. Rev. Lett.},
  volume = {106},
  number = {16},
  pages = {160601},
  publisher = {American Physical Society},
  doi = {10.1103/PhysRevLett.106.160601},
}

@article{evans_stochastic_2020,
  title = {Stochastic Resetting and Applications},
  author = {Evans, M. R. and Majumdar, S. N. and Schehr, G.},
  year = 2020,
  journal = {J. Phys. A: Math. Theor.},
  volume = {53},
  number = {19},
  pages = {193001},
  publisher = {IOP Publishing},
  doi = {10.1088/1751-8121/ab7cfe},
}

@article{falcao_universal_2021,
  title = {Universal Dynamical Scaling Laws in Three-State Quantum Walks},
  author = {Falc{\~a}o, P. R. N. and Buarque, A. R. C. and Dias, W. S. and Almeida, G. M. A. and Lyra, M. L.},
  year = 2021,
  journal = {Phys. Rev. E},
  volume = {104},
  number = {5},
  pages = {054106},
  doi = {10.1103/PhysRevE.104.054106},
  langid = {english},
}

@article{falkner_weak_2014,
  title = {Weak Limit of the Three-State Quantum Walk on the Line},
  author = {Falkner, S. and Boettcher, S.},
  year = 2014,
  journal = {Phys. Rev. A},
  volume = {90},
  number = {1},
  pages = {012307},
  doi = {10.1103/PhysRevA.90.012307},
  langid = {english},
}

@book{feller_introduction_2009,
  title = {An {{Introduction}} to {{Probability Theory}} and Its {{Applications}}. {{Vol}}. 1 and 2},
  author = {Feller, W.},
  year = 2009,
  publisher = {Wiley},
  address = {New York},
  langid = {english}
}

@article{friedman_quantum_2017,
  title = {Quantum Walks: {{The}} First Detected Passage Time Problem},
  shorttitle = {Quantum Walks},
  author = {Friedman, H. and Kessler, D. A. and Barkai, E.},
  year = 2017,
  journal = {Phys. Rev. E},
  volume = {95},
  number = {3},
  pages = {032141},
  doi = {10.1103/PhysRevE.95.032141},
  langid = {english},
}

@article{gotta_towers_2026,
  title = {Towers of Quantum Many-Body Scars under Stochastic Resetting},
  author = {Gotta, L. and Kulkarni, M. and Perfetto, G.},
  year = 2026,
  journal = {arXiv:2603.13165},
  eprint = {2603.13165},
  primaryclass = {cond-mat.stat-mech},
  publisher = {arXiv},
  doi = {10.48550/arXiv.2603.13165},
  archiveprefix = {arXiv},
}

@article{inui_one-dimensional_2005,
  title = {One-Dimensional Three-State Quantum Walk},
  author = {Inui, N. and Konno, N. and Segawa, E.},
  year = 2005,
  journal = {Phys. Rev. E},
  volume = {72},
  number = {5},
  pages = {056112},
  doi = {10.1103/PhysRevE.72.056112},
  langid = {english},
}

@article{kempe_quantum_2003,
  title = {Quantum Random Walks: {{An}} Introductory Overview},
  shorttitle = {Quantum Random Walks},
  author = {Kempe, J.},
  year = 2003,
  journal = {Contemporary Physics},
  volume = {44},
  number = {4},
  pages = {307--327},
  doi = {10.1080/00107151031000110776},
  langid = {english},
}

@article{kiumi_spectral_2026,
  title = {Spectral Analysis of Three-State Quantum Walks with General Coin Matrices},
  author = {Kiumi, C. and Akahori, J. and Watanabe, T. and Konno, N.},
  year = 2026,
  journal = {Phys. Rev. Res.},
  volume = {8},
  number = {2},
  pages = {023201},
  doi = {10.1103/w8q6-zh9q},
  langid = {english},
}

@article{kulkarni_generating_2023,
  title = {Generating Entanglement by Quantum Resetting},
  author = {Kulkarni, M. and Majumdar, S. N.},
  year = 2023,
  journal = {Phys. Rev. A},
  volume = {108},
  number = {6},
  pages = {062210},
  doi = {10.1103/PhysRevA.108.062210},
  langid = {english},
}

@article{lapeyre_unified_2024,
  title = {Unified Approach to Reset Processes and Application to Coupling between Process and Reset},
  author = {Lapeyre, G. J. and Aquino, T. and Dentz, M.},
  year = 2024,
  journal = {Phys. Rev. E},
  volume = {110},
  number = {4},
  pages = {044138},
  doi = {10.1103/PhysRevE.110.044138},
  langid = {english}
}

@article{leykam_artificial_2018,
  title = {Artificial Flat Band Systems: From Lattice Models to Experiments},
  shorttitle = {Artificial Flat Band Systems},
  author = {Leykam, D. and Andreanov, A. and Flach, S.},
  year = 2018,
  journal = {Adv. Phys.: X},
  volume = {3},
  number = {1},
  pages = {1473052},
  doi = {10.1080/23746149.2018.1473052},
  langid = {english},
}

@article{masoliver_telegraphic_2019,
  title = {Telegraphic Processes with Stochastic Resetting},
  author = {Masoliver, J.},
  year = 2019,
  journal = {Phys. Rev. E},
  volume = {99},
  number = {1},
  pages = {012121},
  doi = {10.1103/PhysRevE.99.012121},
  langid = {english},
}

@article{mukherjee_quantum_2018,
  title = {Quantum Dynamics with Stochastic Reset},
  author = {Mukherjee, B. and Sengupta, K. and Majumdar, S. N.},
  year = 2018,
  journal = {Phys. Rev. B},
  volume = {98},
  number = {10},
  pages = {104309},
  publisher = {American Physical Society},
  doi = {10.1103/PhysRevB.98.104309},
}

@article{murauer_nonequilibrium_2026,
  title = {Nonequilibrium Steady States Induced by Stochastic Mid-Circuit Measurements and Resets on a Quantum Computer},
  author = {Murauer, J. and Tornow, S. and Perfetto, G.},
  year = 2026,
  journal = {arXiv:2606.19027},
  eprint = {2606.19027},
  primaryclass = {quant-ph},
  publisher = {arXiv},
  doi = {10.48550/arXiv.2606.19027},
  archiveprefix = {arXiv},
}

@article{nagar_diffusion_2016,
  title = {Diffusion with Stochastic Resetting at Power-Law Times},
  author = {Nagar, A. and Gupta, S.},
  year = 2016,
  journal = {Phys. Rev. E},
  volume = {93},
  number = {6},
  pages = {060102},
  publisher = {American Physical Society},
  doi = {10.1103/PhysRevE.93.060102},
}

@article{navascues_resetting_2018,
  title = {Resetting Uncontrolled Quantum Systems},
  author = {Navascu{\'e}s, M.},
  year = 2018,
  journal = {Phys. Rev. X},
  volume = {8},
  number = {3},
  pages = {031008},
  doi = {10.1103/PhysRevX.8.031008},
  langid = {english},
}

@article{nayak_quantum_2000,
  title = {Quantum Walk on the Line},
  author = {Nayak, A. and Vishwanath, A.},
  year = 2000,
  journal = {arXiv:quant-ph/0010117},
  eprint = {quant-ph/0010117},
  archiveprefix = {arXiv},
}

@article{nitsche_probing_2018,
  title = {Probing Measurement-Induced Effects in Quantum Walks via Recurrence},
  author = {Nitsche, T. and Barkhofen, S. and Kruse, R. and Sansoni, L. and {\v S}tefa{\v n}{\'a}k, M. and G{\'a}bris, A. and Poto{\v c}ek, V. and Kiss, T. and Jex, I. and Silberhorn, C.},
  year = 2018,
  journal = {Sci. Adv.},
  volume = {4},
  number = {6},
  pages = {eaar6444},
  publisher = {American Association for the Advancement of Science},
  doi = {10.1126/sciadv.aar6444},
}

@article{pal_first_2017,
  title = {First Passage under Restart},
  author = {Pal, A. and Reuveni, S.},
  year = 2017,
  journal = {Phys. Rev. Lett.},
  volume = {118},
  number = {3},
  pages = {030603},
  publisher = {American Physical Society},
  doi = {10.1103/PhysRevLett.118.030603},
}

@article{perfetto_designing_2021,
  title = {Designing Nonequilibrium States of Quantum Matter through Stochastic Resetting},
  author = {Perfetto, G. and Carollo, F. and Magoni, M. and Lesanovsky, I.},
  year = 2021,
  journal = {Phys. Rev. B},
  volume = {104},
  number = {18},
  pages = {L180302},
  doi = {10.1103/PhysRevB.104.L180302},
  langid = {english},
}

@article{perfetto_thermodynamics_2022,
  title = {Thermodynamics of Quantum-Jump Trajectories of Open Quantum Systems Subject to Stochastic Resetting},
  author = {Perfetto, G. and Carollo, F. and Lesanovsky, I.},
  year = 2022,
  journal = {SciPost Phys.},
  volume = {13},
  number = {4},
  pages = {079},
  doi = {10.21468/SciPostPhys.13.4.079},
}

@book{portugal_quantum_2013,
  title = {Quantum Walks and Search Algorithms},
  author = {Portugal, R.},
  year = 2013,
  publisher = {Springer},
  address = {New York},
  doi = {10.1007/978-1-4614-6336-8},
  langid = {english}
}

@article{qiao_emergence_2026,
  title = {Emergence of Drifted Diffusion in Quantum Walks with Subspace Restart},
  author = {Qiao, L. and Yin, R. and Zhang, W.},
  year = 2026,
  journal = {arXiv:2607.12727},
  eprint = {2607.12727},
  publisher = {arXiv},
  doi = {10.48550/ARXIV.2607.12727},
  archiveprefix = {arXiv},
}

@article{reuveni_optimal_2016,
  title = {Optimal Stochastic Restart Renders Fluctuations in First Passage Times Universal},
  author = {Reuveni, S.},
  year = 2016,
  journal = {Phys. Rev. Lett.},
  volume = {116},
  number = {17},
  pages = {170601},
  publisher = {American Physical Society},
  doi = {10.1103/PhysRevLett.116.170601},
}

@article{roy_causality_2025,
  title = {Causality, Localization, and Universality of Monitored Quantum Walks with Long-Range Hopping},
  author = {Roy, S. and Gupta, S. and Morigi, G.},
  year = 2025,
  journal = {Phys. Rev. E},
  volume = {112},
  number = {4},
  pages = {044146},
  doi = {10.1103/rbtb-8d27},
  langid = {english},
}

@article{schreiber_photons_2010,
  title = {Photons Walking the {{Line}}: {{A}} Quantum Walk with Adjustable Coin Operations},
  shorttitle = {Photons {{Walking}} the {{Line}}},
  author = {Schreiber, A. and Cassemiro, K. N. and Poto{\v c}ek, V. and G{\'a}bris, A. and Mosley, P. J. and Andersson, E. and Jex, I. and Silberhorn, {\relax Ch}.},
  year = 2010,
  journal = {Phys. Rev. Lett.},
  volume = {104},
  number = {5},
  pages = {050502},
  doi = {10.1103/PhysRevLett.104.050502},
  langid = {english},
}

@article{shenvi_quantum_2003,
  title = {Quantum Random-Walk Search Algorithm},
  author = {Shenvi, N. and Kempe, J. and Whaley, K. B.},
  year = 2003,
  journal = {Phys. Rev. A},
  volume = {67},
  number = {5},
  pages = {052307},
  doi = {10.1103/PhysRevA.67.052307},
  langid = {english},
}

@article{shukla_accelerated_2025,
  title = {Accelerated First Detection in Discrete-Time Quantum Walks Using Sharp Restarts},
  author = {Shukla, K. and Chatterjee, R. and Chandrashekar, C. M.},
  year = 2025,
  journal = {Phys. Rev. Res.},
  volume = {7},
  number = {2},
  pages = {023069},
  publisher = {American Physical Society},
  doi = {10.1103/PhysRevResearch.7.023069},
}

@article{singh_general_2022,
  title = {General Approach to Stochastic Resetting},
  author = {Singh, R. K. and G{\'o}rska, K. and Sandev, T.},
  year = 2022,
  journal = {Phys. Rev. E},
  volume = {105},
  number = {6},
  pages = {064133},
  doi = {10.1103/PhysRevE.105.064133},
  langid = {english},
}

@article{smith_renewal_1958,
  title = {Renewal Theory and Its Ramifications},
  author = {Smith, W. L.},
  year = 1958,
  journal = {J. R. Stat. Soc. B (Method.)},
  volume = {20},
  number = {2},
  pages = {243--284},
  doi = {10.1111/j.2517-6161.1958.tb00294.x},
  langid = {english}
}

@article{stefanak_continuous_2012,
  title = {Continuous Deformations of the {{Grover}} Walk Preserving Localization},
  author = {{\v S}tefa{\v n}{\'a}k, M. and Bezd{\v e}kov{\'a}, I. and Jex, I.},
  year = 2012,
  journal = {Eur. Phys. J. D},
  volume = {66},
  number = {5},
  pages = {142},
  doi = {10.1140/epjd/e2012-30146-9},
  langid = {english},
}

@article{stefanak_limit_2014,
  title = {Limit Distributions of Three-State Quantum Walks: {{The}} Role of Coin Eigenstates},
  shorttitle = {Limit Distributions of Three-State Quantum Walks},
  author = {{\v S}tefa{\v n}{\'a}k, M. and Bezd{\v e}kov{\'a}, I. and Jex, I.},
  year = 2014,
  journal = {Phys. Rev. A},
  volume = {90},
  number = {1},
  pages = {012342},
  doi = {10.1103/PhysRevA.90.012342},
  langid = {english},
}

@article{stefanak_monitored_2023,
  title = {Monitored Recurrence of a One-Parameter Family of Three-State Quantum Walks},
  author = {{\v S}tefa{\v n}{\'a}k, M.},
  year = 2023,
  journal = {Phys. Scr.},
  volume = {98},
  number = {6},
  pages = {064001},
  doi = {10.1088/1402-4896/accf43},
}

@article{thiel_first_2018,
  title = {First Detected Arrival of a Quantum Walker on an Infinite Line},
  author = {Thiel, F. and Barkai, E. and Kessler, D. A.},
  year = 2018,
  journal = {Phys. Rev. Lett.},
  volume = {120},
  number = {4},
  pages = {040502},
  doi = {10.1103/PhysRevLett.120.040502},
  langid = {english},
}

@article{wald_classical_2021,
  title = {From Classical to Quantum Walks with Stochastic Resetting on Networks},
  author = {Wald, S. and B{\"o}ttcher, L.},
  year = 2021,
  journal = {Phys. Rev. E},
  volume = {103},
  number = {1},
  pages = {012122},
  publisher = {American Physical Society},
  doi = {10.1103/PhysRevE.103.012122},
}

@article{wald_stochastic_2025,
  title = {Stochastic Resetting in Discrete-Time Quantum Dynamics: Steady States and Correlations in Few-Qubit Systems},
  shorttitle = {Stochastic Resetting in Discrete-Time Quantum Dynamics},
  author = {Wald, S. and Yao, L. H. and Platini, T. and Hooley, C. and Carollo, F.},
  year = 2025,
  journal = {Quantum},
  volume = {9},
  pages = {1742},
  doi = {10.22331/q-2025-05-13-1742},
  langid = {english}
}

@article{wang_first_2024,
  title = {First Hitting Times on a Quantum Computer: {{Tracking}} vs. Local Monitoring, Topological Effects, and Dark States},
  shorttitle = {First {{Hitting Times}} on a {{Quantum Computer}}},
  author = {Wang, Q. and Ren, S. and Yin, R. and Ziegler, K. and Barkai, E. and Tornow, S.},
  year = 2024,
  journal = {Entropy},
  volume = {26},
  number = {10},
  pages = {869},
  doi = {10.3390/e26100869},
  langid = {english}
}

@article{wang_one-dimensional_2017,
  title = {One-Dimensional Lackadaisical Quantum Walks},
  author = {Wang, K. and Wu, N. and Xu, P. and Song, F.},
  year = 2017,
  journal = {J. Phys. A: Math. Theor.},
  volume = {50},
  number = {50},
  pages = {505303},
  doi = {10.1088/1751-8121/aa9235},
}

@article{wojcik_re-examining_2026,
  title = {Re-Examining Quantum Walk Advantages: {{A}} Mean Hitting Time Perspective},
  shorttitle = {Re-Examining Quantum Walk Advantages},
  author = {W{\'o}jcik, J.},
  year = 2026,
  journal = {Phys. Rev. E},
  volume = {113},
  number = {2},
  pages = {024135},
  doi = {10.1103/5982-n6c3},
  langid = {english},
}

@article{wong_coined_2017,
  title = {Coined Quantum Walks on Weighted Graphs},
  author = {Wong, T. G.},
  year = 2017,
  journal = {J. Phys. A: Math. Theor.},
  volume = {50},
  number = {47},
  pages = {475301},
  doi = {10.1088/1751-8121/aa8c17},
}

@article{wong_grover_2015,
  title = {Grover Search with Lackadaisical Quantum Walks},
  author = {Wong, T. G.},
  year = 2015,
  journal = {J. Phys. A: Math. Theor.},
  volume = {48},
  number = {43},
  pages = {435304},
  doi = {10.1088/1751-8113/48/43/435304},
}

@article{yin_instability_2024,
  title = {Instability in the Quantum Restart Problem},
  author = {Yin, R. and Wang, Q. and Barkai, E.},
  year = 2024,
  journal = {Phys. Rev. E},
  volume = {109},
  number = {6},
  pages = {064150},
  doi = {10.1103/PhysRevE.109.064150},
  langid = {english}
}

@article{yin_restart_2023,
  title = {Restart Expedites Quantum Walk Hitting Times},
  author = {Yin, R. and Barkai, E.},
  year = 2023,
  journal = {Phys. Rev. Lett.},
  volume = {130},
  number = {5},
  pages = {050802},
  publisher = {American Physical Society},
  doi = {10.1103/PhysRevLett.130.050802},
}
\end{document}